\documentclass[prx,superscriptaddress,twocolumn,floats,10pt,floatfix,nobibnotes]{revtex4-1}

\usepackage{amsmath}
\usepackage{amsfonts}
\usepackage{amssymb}
\usepackage{graphicx}
\usepackage{color}
\usepackage[colorlinks=true,citecolor=blue,linkcolor=blue,urlcolor=black]{hyperref}
\usepackage{times}
\usepackage{amstext}
\usepackage{latexsym}
\usepackage{bbm}

\begin{document}

\title{Spectral Classification of Coupling Regimes in the Quantum Rabi Model}
\author{Daniel Z. Rossatto}
\email{zini@df.ufscar.br}
\affiliation{Departamento de F\'{\i}sica, Universidade Federal de S\~{a}o Carlos, 13565-905, S\~{a}o Carlos, SP, Brazil}
\author{Celso J. Villas-B\^{o}as}
\affiliation{Departamento de F\'{\i}sica, Universidade Federal de S\~{a}o Carlos, 13565-905, S\~{a}o Carlos, SP, Brazil}
\author{Mikel Sanz}
\email{mikel.sanz@ehu.eus}
\affiliation{Department of Physical Chemistry, University of the Basque Country UPV/EHU, Apartado 644, E-48080 Bilbao, Spain}
\author{ Enrique Solano}
\affiliation{Department of Physical Chemistry, University of the Basque Country UPV/EHU, Apartado 644, E-48080 Bilbao, Spain}
\affiliation{IKERBASQUE, Basque Foundation for Science, Mar\'{i}a D\'{i}az de Haro 3, E-48013 Bilbao, Spain}

\begin{abstract}
The quantum Rabi model is in the scientific spotlight due to the recent theoretical and experimental progress. Nevertheless, a full-fledged classification of its coupling regimes remains as a relevant open question. We propose a spectral classification dividing the coupling regimes into three regions based on the validity of perturbative criteria on the quantum Rabi model, which allows us the use of exactly solvable effective Hamiltonians. These coupling regimes are i) the perturbative ultrastrong coupling regime which comprises the Jaynes-Cummings model, ii) a region where non-perturbative ultrastrong and non-perturbative deep strong coupling regimes coexist, and  iii) the perturbative deep strong coupling regime. We show that this spectral classification depends not only on the ratio between the coupling strength and the natural frequencies of the unperturbed parts, but also on the energy to which the system can access. These regimes additionally discriminate the completely different behaviors of several static physical properties, namely the total number of excitations, the photon statistics of the field, and the cavity-qubit entanglement. Finally, we explain the dynamical properties which are traditionally associated to the deep strong coupling regime, such as the collapses and revivals of the state population, in the frame of the proposed spectral classification. 
\end{abstract}

\maketitle

\section{Introduction}
\label{sec:introd}
The well-established Rabi model \cite{rabi1936} describes the simplest class of light-matter interaction, the dipolar coupling between a two-level quantum system (qubit) and a classical monochromatic radiation field (unidimensional harmonic oscillator). In its quantum version, the radiation is specified by a quantized single-mode field, yielding the so-called quantum Rabi model (QRM) \cite{braak2011, solano2011}. This model accurately describes the dynamics of a wide variety of physical setups, ranging from quantum optics to condensed matter systems \cite{book2013}. In addition, a plethora of protocols in contemporary quantum information theory \cite{nielsen2004}, with potential applications in future quantum technologies covering from ultrafast gates \cite{romero2012} to quantum error correcting codes \cite{kyawprb2015} or remote entanglement generation \cite{felicettiprl2014,rossatto2016}, make use of the QRM as a building block. Therefore, the QRM plays an extremely important role in both theoretical and applied physics.

Typically, the standard experiments on cavity quantum electrodynamics (cavity QED) are restricted to a light-matter coupling strength much smaller than the natural frequencies of the unperturbed parts. Thus, they happen in the realm of the renowned Jaynes-Cummings (JC) model \cite{jc1963}, which is obtained by applying the rotating-wave approximation (RWA) to the QRM \cite{knight2005}. In this scenario, the achievement of the so-called strong coupling (SC) regime, when the coupling strength exceeds all decoherence rates, has driven the field of cavity QED for several decades \cite{book2013}. Therefore, the JC model has served as a theoretical and experimental milestone in the history of quantum physics.

Since the last decade, a new coupling regime of the QRM, in which the coupling strength is a substantial fraction of the natural frequencies of the unperturbed parts, is being theoretically studied \cite{ciuti2005,bourassa2009,beaudoin2011,ballester2012,pedernales2015, todorov2010} and experimentally reached in diverse solid state systems \cite{todorov2010, anappara2009, gunter2009, forndiaz2010, niemczyk2010, federov2010, muravev2011, schwartz2011, scalari2012, geiser2012, goryachev2014, zhang2016, chen2016, langford2016, braumueller2016}. In this so-called ultrastrong coupling (USC) regime, the RWA is no longer suitable, such that the counter-rotating terms provide novel counterintuitive physical phenomena and new applications for the QRM emerge \cite{ashab2010, nataf2012, ridolfo2012, romero2012, ridolfo2013, stassi2013, felicetti2014, garziano2014,kyawprb2015, felicettiprl2014, garziano2015, kyaw2015, felicetti2015, rossatto2016, forndiaz2016, wang2016}. When the counter-rotating terms can still be perturbatively treated, as in Refs.~\cite{todorov2010, anappara2009, gunter2009, forndiaz2010, niemczyk2010, federov2010, muravev2011, schwartz2011, scalari2012, geiser2012, goryachev2014, zhang2016, chen2016}, the QRM is approximately described by the Bloch-Siegert (BS) Hamiltonian \cite{forndiaz2010, beaudoin2011,klimovbook}. However, a few experiments have recently achieved the non-perturbative USC regime \cite{maissen2014, gambino2014, forndiaz2016-2, yoshihara2016,yoshihara2016-2}, for which the full QRM has to be considered.

When the coupling strength is even stronger, surpassing the natural frequencies of the unperturbed parts, another regime of light-matter interaction appears, with totally different physics than the USC regime \cite{casanova2010, liberato2014}. For this so-called deep strong coupling (DSC) regime \cite{casanova2010}, the QRM can be reasonably described by an approximate solution as discussed in Refs.~\cite{casanova2010,feranchuk1996, irish2005, irish2007}. And, recently, F.~Yoshihara \textit{et al}.~have experimentally achieved such an impressive coupling in superconducting circuits \cite{yoshihara2016,yoshihara2016-2}.

Therefore, the advent of the aforementioned remarkable experimental and theoretical achievements has placed the QRM in the scientific spotlight. Nonetheless, the characterization so far established for the coupling regimes of the QRM is not quite universal, and a more specific criterion still remains undetermined. For instance, there are definitions stating that the USC regime is reached when the coupling strength is greater than a critical value related to either dynamical correlation functions \cite{wolf2013} or quantum phase transitions \cite{domokos2015,jaako2016}. However, for the latter case there is no consensus whether this transition can be reached in physical setups  \cite{jaako2016,nataf2010,viehmann2011,ciuti2012,bamba2016}, and this definition does not take into account the properties of the whole model, but only of its ground state. Another attempt was recently proposed in Ref.~\cite{yoshihara2016-2}, where the coupling classification is based on unique features exhibited in the transmission spectra of the system for different coupling regions. This approach uses the fact that the selection rules which allow or forbid transitions between eigenstates depend on the coupling value, changing the transmission pattern for different coupling regions. However, similarly to Refs.~\cite{domokos2015,jaako2016}, this approach does not take into account the properties of the whole model, since Another attempt was recently proposed in Ref.~\cite{yoshihara2016-2}, where the coupling classification is based on unique features exhibited in the transmission spectra of the system for different coupling regions. This approach uses the fact that the selection rules which allow or forbid transitions between eigenstates depend on the coupling value, changing the transmission pattern for different coupling regions. However, similarly to Refs.~\cite{domokos2015,jaako2016}, this approach does not take into account the properties of the whole model, since only the first four eigenstates are considered.

On the other hand, it is also conjectured in the literature that the USC regime is achieved when the coupling strength is just a substantial fraction of the natural frequencies of the unperturbed systems \cite{bourassa2009, todorov2010, forndiaz2010, niemczyk2010, scalari2012}. Here, we are interested in quantitatively establishing how substantial this fraction has to be for the system description being significantly affected by the counter-rotating terms. Although the exact analytical solution of the QRM was recently presented for all parameter regimes \cite{braak2011}, it strongly depends on zeros of a transcendental function defined through an infinite power series, making it difficult to extract the fundamental physics of that solution in general. Hence, it is more convenient to use approximate versions of the QRM as far as possible.

In this paper, we show that these approximate solutions are excellent guides to define a quantitative characterization of the coupling regimes of the QRM. In Sec.~\ref{sec:coupreg}, we show that the coupling regimes are naturally divided into three regions, whose boundaries depend not only on the ratio between the coupling strength and the natural frequencies of the unperturbed parts, but also on the energy to which the system can access. In addition, we show in Sec.~\ref{sec:static} that our classification is supported by a completely different behavior of several static physical properties of the QRM, which depends on the region. Section~\ref{sec:dynam} provides a connection of our spectral classification with the dynamical properties that yield the traditionally blurry transition between the USC/DSC regimes. Finally, Section~\ref{sec:conc} comprises the conclusions of our work and the novel open questions emerging from it.

\section{Coupling regimes of the quantum Rabi model}
\label{sec:coupreg}

The Hamiltonian of the ubiquitous QRM is ($\hbar = 1$)
\begin{equation} \label{qrm}
H_{R} = \omega a^{\dag}a +  \frac{\Omega}{2}\sigma_{z} + g_{0}\sigma_{x}(a+ a^{\dag}).
\end{equation}
Here, $\sigma_{x,y,z}$ are the Pauli matrices for the qubit, with transition frequency $\Omega$ ($\vert \text{g} \rangle =$ ground state and $\vert \text{e} \rangle =$ excited state), and $a$ ($a^{\dagger}$) stands for the annihilation (creation) operator of a single-mode bosonic field, with frequency $\omega$. The light-matter coupling is quantified by the vacuum-Rabi frequency $g_{0}$.

\subsection{Perturbative ultrastrong coupling regime}
\label{sec:pUSCdef}

Whenever $\vert \delta \vert \ll g_{0}\sqrt{\langle \hat{n}\rangle +1} \ll \Sigma$, with $\delta = \Omega - \omega$, $\Sigma = \Omega + \omega$, and $\langle \hat{n}\rangle=\langle a^{\dagger}a\rangle$, the QRM is well described by the JC model using the RWA \cite{knight2005},
\begin{equation} \label{jcm}
H_{\text{JC}} = \omega a^{\dag}a +  \frac{\Omega}{2}\sigma_{z} + g_{0}(a\sigma_{+}+ a^{\dag}\sigma_{-}),
\end{equation}
with $\sigma_{\pm} = (\sigma_{x} \pm i\sigma_{y})/2$. Paradigmatic examples of the intuitive physics behind the JC dynamics are the Rabi oscillations in the JC doublets, Eqs.~\eqref{pnBS} and \eqref{mnBS} with $\omega_{\text{BS}} = g_{0}^{2}/\Sigma \rightarrow 0$, as a consequence of the conservation of the total number of excitations, and the collapses and revivals of the population inversion of the qubit \cite{knight2005}.

When the counter-rotating terms can still be perturbatively treated, it is convenient to use the unitary transformation
\begin{equation} \label{unitrans}
\mathcal{U} = \exp{\textbf{(} \Lambda(a \sigma_{-} - a^{\dagger}\sigma_{+}) + \xi(a^{2} - a^{\dagger 2})\sigma_{z}\textbf{)}},
\end{equation}
with $\Lambda = g_{0}/(\omega + \Omega)$ and $\xi = g_{0}\Lambda / 2\omega$. To second order in $\Lambda$, this yields the Bloch-Siegert (BS) Hamiltonian \cite{forndiaz2010,beaudoin2011,klimovbook}
%
\begin{equation} 
\mathcal{U}^{\dagger}H_{R}\mathcal{U} \approx H_{\text{BS}}^{(2)} = \omega_{\text{BS}} \sigma_{z}a^{\dag}a + \omega_{\text{BS}} \frac{\sigma_{z}}{2} - \frac{\omega_{\text{BS}}}{2} + H_{\text{JC}}, \label{bs}
\end{equation}
in which $\omega_{\text{BS}} = g_{0}^{2}/\Sigma$ is the BS shift. From Eq.~(\ref{bs}), it is straightforward to note that while the BS Hamiltonian provides the second-order correction in $\Lambda$, the JC provides the zeroth-order one. Hence, the JC model is recovered from the BS Hamiltonian by imposing $\Lambda = 0 \implies \omega_{\text{BS}} = 0$.

 The energy spectrum of the BS Hamiltonian is
\begin{align} 
E_{0}^{\text{BS}} &=  -\frac{\Omega}{2} - \omega_{\text{BS}} \label{EgBS}, \\  
E_{n,\pm}^{\text{BS}} &= \left(n - \frac{1}{2} \right)\omega - \omega_{\text{BS}} \pm \frac{1}{2}\sqrt{(\Delta_{n}^{\text{BS}})^{2} + 4g_{0}^{2}n},  \label{EpmBS}
\end{align}
with $\Delta_{n}^{\text{BS}} = \delta + 2\omega_{\text{BS}}n$ and $n \in \mathbb{N}^{*}$. The eigenstates are 
\begin{align} 
\vert\phi_{0}^{\text{BS}}\rangle &=  \mathcal{U}\vert \text{g},0 \rangle \label{phigBS}, \\  
\vert\phi_{n,\pm}^{\text{BS}} \rangle &= \mathcal{U}\vert \pm,n \rangle  \label{lphiEpmBS},
\end{align}
with $\vert \text{g},n \rangle=\vert \text{g}\rangle \otimes \vert n \rangle$, where $\vert n \rangle$ is the Fock state, and 
\begin{align} 
&\vert +,n \rangle = \cos{\frac{\theta_{n}}{2}}\vert \text{e},n-1 \rangle +\sin{\frac{\theta_{n}}{2}}\vert \text{g},n \rangle  \label{pnBS}, \\  
&\vert -,n \rangle =  \sin{\frac{\theta_{n}}{2}}\vert \text{e},n-1 \rangle - \cos{\frac{\theta_{n}}{2}}\vert \text{g},n \rangle \label{mnBS},
\end{align}
in which the BS mixing angle is
\begin{equation}
\theta_{n} = \arctan{\left ( \frac{2g_{0}\sqrt{n} }{\Delta_{n}^{\text{BS}}} \right )}. \label{BSangle}
\end{equation}

\begin{figure}[t!]
\includegraphics[trim = 10mm 7mm 10mm 13mm, clip, width=0.5\textwidth]{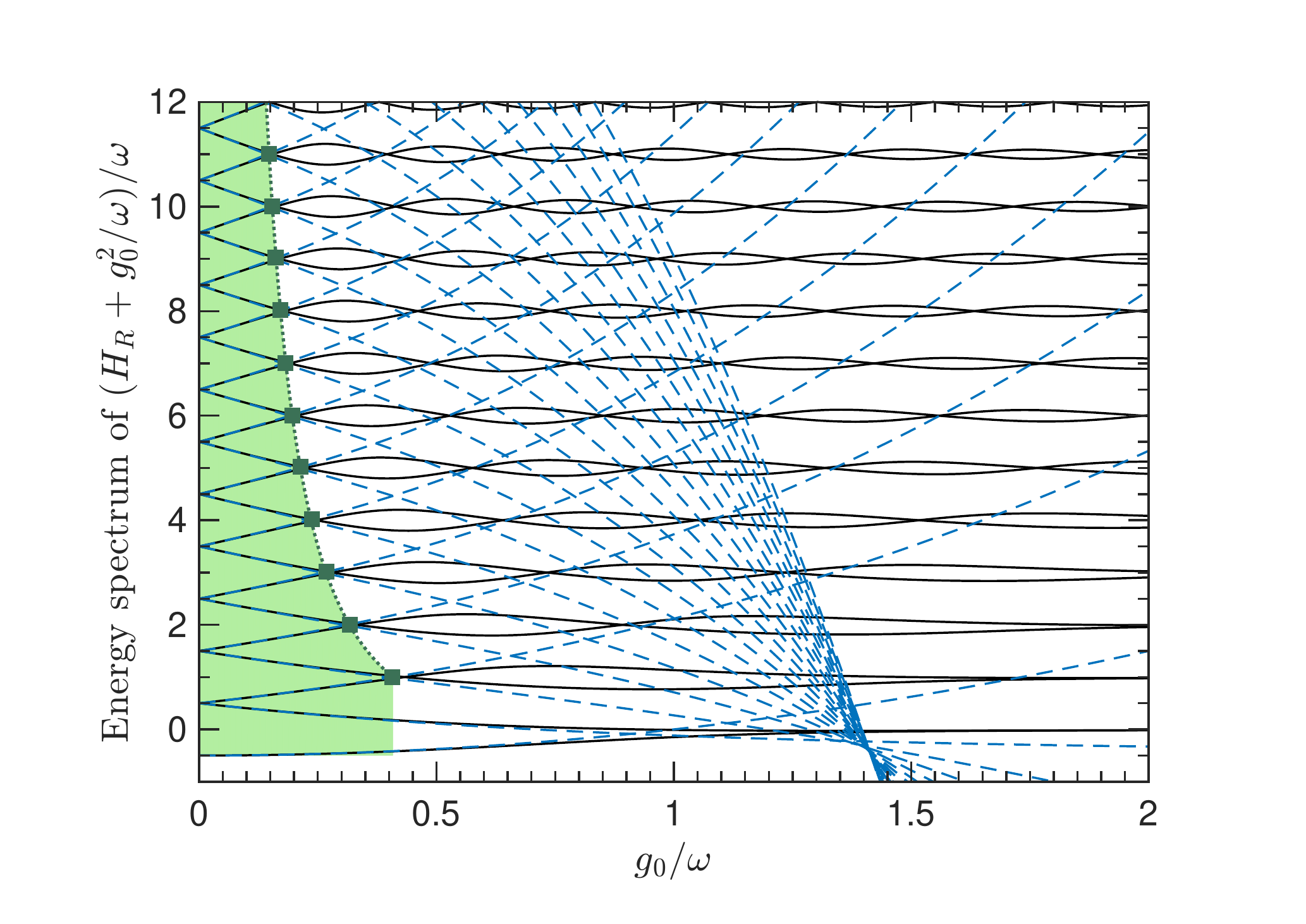}
\caption{Energy spectrum of the QRM (solid lines) and BS energy spectrum (dashed lines) vs $g_{0}/\omega$. The squares represent the first Juddian points calculated using Eq.~(\ref{g_cr_BS}), while the shaded area is the region where the perturbative USC regime is valid [Eq.~\eqref{pUSCregion}]. For the sake of clarity, the eigenenergies are rescaled by $g_{0}^{2}/\omega$.}
\label{fig_bs}
\end{figure}

For the sake of simplicity, we will consider the resonant case  ($\omega=\Omega$) hereafter, but it is worth stressing that the following discussion is also suitable for the general case. In Fig.~\ref{fig_bs}, we observe that the BS Hamiltonian provides an energy spectrum in excellent agreement with the one of the full QRM \cite{fidBS}, surprisingly up to the first energy-level crossings (the so-called Juddian points \cite{braak2011}).

Therefore, the use of the first Juddian points is an excellent attempt to define a boundary for a coupling regime. In this case, since the BS Hamiltonian perturbatively takes into account the breakdown of the RWA, we can establish the \textit{perturbative USC regime} (pUSC) of the QRM as the region before the first Juddian points ($g_{\text{pUSC}}^{\times}$), which is obtained by imposing $E_{n,+}^{\text{BS}}=E_{n+1,-}^{\text{BS}}$. By squaring both sides of $E_{n,+}^{\text{BS}}=E_{n+1,-}^{\text{BS}}$ up to the elimination of the square roots, we end up with an eighth-degree polynomial in $g_{0}/\omega$. Since the BS Hamiltonian is valid for perturbative values of $g_{0}/\omega$, we truncate this polynomial up to second order, so that its non-negative solution yields the first Juddian points for each $n \ge 1$,
\begin{equation} \label{g_cr_BS}
\frac{g_{\text{pUSC}}^{\times}}{\omega} \simeq \frac{1}{\sqrt{2(2n+1)}},
\end{equation}
such that $0 \le g_{\text{pUSC}}^{\times}/\omega \lesssim 1/\sqrt{6}$.

We can also notice from Fig.~\ref{fig_bs} that the more energetic the eigenenergies, the smaller the coupling values of first the Juddian points. This indicates that the importance of the counter-rotating terms depends not only on the ratio $g_{0}/\omega$, but also on the energy to which the system can access, showing that the properties of the ground state are not sufficient to fully classify the coupling regimes of the QRM. Thus, the definition of the pUSC coupling regime is also connected to the energy to which the system can access.

Let us enlarge upon this point for the sake of clarity. The question we want to answer is whether a quantum state $|\psi\rangle$ evolving under $H_R$ with a given $g_0/\omega$ will show features corresponding to the pUSC regime. This state is not necessarily an eigenstate, but it may be expanded in terms of eigenstates of the QRM. Thus, for a given $g_{0}/\omega$, $\vert \psi\rangle$ can have contributions from eigenstates in the pUSC region and from the region beyond that. Therefore, we take as a natural qualitative quantifier the mean energy of the state $\bar{\mathcal{E}} = \langle \psi|H_R|\psi\rangle$ and choose the criterion that this state is in the pUSC regime when its energy is below the curve shown in the following.


If we invert Eq.~(\ref{g_cr_BS}) and replace $n$ in $E_{n,+}^{\text{BS}}$, assuming it as a continuous parameter, we can define the boundary of the pUSC regime as 
\begin{align}
\frac{\mathcal{E}_{\text{pUSC}}}{\omega} &\simeq \frac{1}{4}\left(\frac{\omega}{g_{0}}\right)^{2} \left[ 1 - 2\left( \frac{g_{0}}{\omega} \right)^{4}  \right]- 1 \nonumber \\
&+ \frac{1}{4}\sqrt{\left[ 5 - 2\left( \frac{g_{0}}{\omega} \right)^{2} \right]  \left[ 1 - 2\left( \frac{g_{0}}{\omega} \right)^{2} \right] }. \label{bound_BS}
\end{align}
This boundary is illustrated as the dotted line in Fig.~\ref{fig_bs}, with the shaded area standing for the region where the perturbative USC regime is valid, i.e., when
\begin{equation} \label{pUSCregion}
g_{0} \lesssim g_{\text{pUSC}}^{\times} \quad \text{and} \quad  \bar{\mathcal{E}} \lesssim \mathcal{E}_{\text{pUSC}}.
\end{equation}

It is worth stressing that, besides the BS approach, there are other ones that can describe more accurately the QRM in a perturbative way \cite{feranchuk1996,irish2005, irish2007, jia2012}. However, these methods result in a much more complicated solution for $g_{\text{pUSC}}^{\times}$. We have also checked that the BS Hamiltonian expanded to third order in $g_{0}/\omega$ \cite{forndiaz2016} provides more accurate eigenenergies, which also diverge from the exact calculated ones beyond the first Juddian points (see Appendix A). This indicates that the proposed definition for the pUSC region is not a simple consequence of the second-order term, but something deeper related to the breaking of the assumptions for the adiabatic expansion and the point from which the total number of excitation is no longer preserved, as discussed in Sec.~\hyperref[sec:static]{III} and shown in the left panel of Fig.~\ref{neM_figure}(a).

\subsection{Perturbative deep strong coupling regime}
\label{sec:pDSCdef}

Analogously to the previous case, we can also employ the same ideas for the coupling regime at the other end, i.e., when the interaction term is no more a mere perturbation, but the main driver of the dynamics (DSC regime). In order to visualize the essence of this regime, it is convenient to rewrite $H_{R}$ in terms of the parity operator $\Pi=-\sigma_{z}(-1)^{a^{\dagger}a}$ \cite{casanova2010}, a conserved quantity of the QRM besides the total energy \cite{braak2011},
\begin{equation} \label{qrm_parity}
H_{R} = \omega b^{\dag}b + g_{0}(b+ b^{\dag}) -  \frac{\Omega}{2}(-1)^{b^{\dag}b}\Pi,
\end{equation}
with $b = \sigma_{x}a$. Since $\Pi$ has eigenvalues $p = \pm 1$ and $[\Pi,H_{R}]=0$, there exists an independent Hamiltonian describing a perturbed displaced harmonic oscillator for each parity chain ($p=\pm1$), whose perturbation is given by the qubit term as an energy shift proportional to $\Omega$ \cite{casanova2010}. Thus, a perturbative approach up to first order in $\Omega$ provides the following energy spectrum and the zeroth-order eigenstates (adiabatic approximation) \cite{irish2005,feranchuk1996,casanova2010}
\begin{align} 
E_{n,\pm}^{\text{pDSC}} &=  (n - \alpha^{2})\omega \pm \frac{\Omega}{2}e^{-2\alpha^{2}} L_{n}(4\alpha^{2}), \label{EpmDSCp} \\
\vert \phi^{\text{pDSC}}_{\pm,n} \rangle &=\frac{1}{\sqrt{2}} \left[ \vert + \rangle \otimes \mathcal{D}(-\alpha) \vert n \rangle  \pm \vert - \rangle \otimes \mathcal{D}(\alpha) \vert n \rangle \right], \label{DSCsol}
\end{align}
in which $ L_{n}(x)$ is the Laguerre polynomial, $\vert \pm \rangle = (\vert \text{e} \rangle \pm \vert \text{g} \rangle)/\sqrt{2}$, $\mathcal{D}(\alpha) = e^{\alpha a^{\dagger} - \alpha^{*}a}$ with $\alpha = g_{0}/\omega$, and $n \in \mathbb{N}$. The energy of the ground state for this case is given by $E_{0,-}^{\text{pDSC}}$. A more refined approximation improves only marginally the accuracy of the eigenenergies and does not reveal new physical behavior \cite{wolf2013}.

\begin{figure}[t]
\includegraphics[trim = 10mm 7mm 10mm 13mm, clip, width=0.5\textwidth]{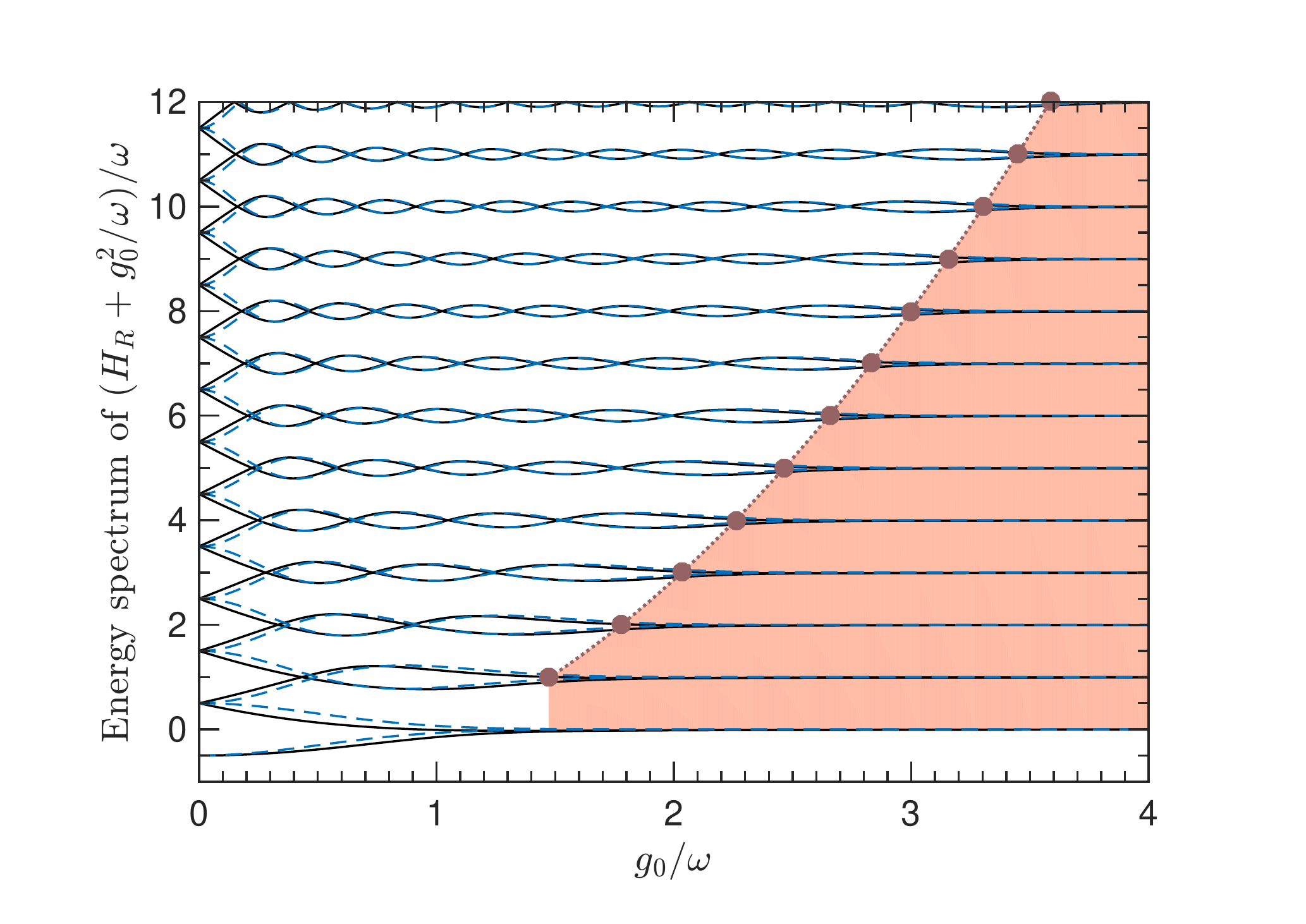}
\caption{Comparison between the energy spectrum of the QRM (solid lines) and the adiabatic approximation (dashed lines) given by Eq.~\eqref{EpmDSCp}, with the circles representing the first solutions of Eq.~\eqref{transeq} for $\delta=0.1$ (Table \ref{pDSCpoints}). The shaded area is the region where the perturbative DSC regime is valid [Eq.~\eqref{pDSCregion}]. For the sake of clarity, the eigenenergies are rescaled by $g_{0}^{2}/\omega$.}
\label{fig_dsc}
\end{figure}

The DSC regime has a typical dynamical feature, which is the appearance of photon-number wave packets that bounce back and forth along a defined parity chain, yielding collapses and revivals of the initial population, even when the field and the qubit are initially in the vacuum and ground state, respectively \cite{casanova2010}. This feature appears only for sufficient large values of $g_{0}/\omega$, and it is more prominent after the last Juddian points (last energy-level crossings), when the adjacent eigenenergies asymptotically approach, becoming quasi-degenerate. Notice that the spectrum and the eigenstates of the QRM are described by Eqs.~\eqref{EpmDSCp} and \eqref{DSCsol} with high fidelity \cite{wolf2013}. Therefore, it is straightforward to note that the collapse-revival phenomenon is strictly related to the Schr\"{o}dinger-cat-like states given by Eq.~\eqref{DSCsol}, which makes the perturbative solution an excellent attempt to define a boundary for a coupling regime.

In Fig.~\ref{fig_dsc}, considering $\omega=\Omega$, we notice that the energy spectrum given by Eq.~\eqref{EpmDSCp} strongly agrees with the one of the full QRM beyond the last Juddian points, when the adjacent eigenenergies become quasi-degenerate. Thus, we use this fact to establish the boundary delimiting the \textit{perturbative DSC} (pDSC) region. The boundary also connects with the appearance of the collapse and revivals of the initial population. For this purpose, let us consider the set of $\alpha_k(n,\delta)$, with $n \ge 1$ and $k=1,2,\dots,m \le n$, which are solutions of the equation
\begin{equation}\label{transeq}
\frac{1}{\omega} \left| E_{+,n}^{\text{pDSC}}- E_{-,n}^{\text{pDSC}}\right|= e^{-2 \alpha^2} \left | L_n(4\alpha^2)\right| \equiv \delta,
\end{equation}
where $\delta$ is the maximum allowed energy difference close to the quasi-degenerate-energy region, which is related to the minimum fidelity allowed between the exact solution of the QRM and the perturbative states  $|\phi_{\pm,n}^{\text{pDSC}}\rangle$. Therefore, the limit of the pDSC region is given by the set of largest solutions of the above transcendental equation, $\{g_{\text{pDSC}}^{\times}/\omega= \max [\alpha_k(n,\delta)]\}$.

For our calculations, we have chosen $\delta =0.1$, value for which we have numerically observed better than $99\%$ fidelity between $\vert \phi^{\text{pDSC}}_{\pm,n} \rangle$ and the corresponding exact eigenstates of the QRM. The numerical solutions of Eq.~\eqref{transeq} corresponding to the lowest values of $n$ are provided in Table~\ref{pDSCpoints}. Inserting $g_{\text{pDSC}}^{\times}$ into  $(E_{+,n}^{\text{pDSC}}+ E_{-,n}^{\text{pDSC}})/2$, these points can be accurately fitted by the second-order equation 
\begin{equation} \label{Epdsc}
\frac{\mathcal{E}_{\text{pDSC}}}{\omega} + \left(\frac{g_{0}}{\omega}\right)^{2} \simeq a\left(\frac{g_{0}}{\omega} \right)^2+b \left(\frac{g_{0}}{\omega} \right) + c,
\end{equation}
with $a = 1.0425$, $b = - 0.054478$ and $c = - 1.1987$, which is our definition of the boundary of the pDSC regime. Such a boundary is illustrated as the dotted line in Fig.~\ref{fig_dsc}, with the shaded area indicating the region where the perturbative DSC regime is valid, i.e., when
\begin{equation} \label{pDSCregion}
g_{0} \gtrsim g_{\text{pDSC}}^{\times} \quad \text{and} \quad  \bar{\mathcal{E}} \lesssim \mathcal{E}_{\text{pDSC}}.
\end{equation}

\begin{table}
\caption{\label{pDSCpoints} First numerical solutions of Eq.~\eqref{transeq} for a $\delta = 0.1$.}
\begin{ruledtabular}
\begin{tabular}{cccc}
$n$  &  $g_{\text{pDSC}}^{\times}/\omega$  &  $n$ &  $g_{\text{pDSC}}^{\times}/\omega$\\
\hline
$1$  &  $1.473$   &  $7$  &   $2.832$ \\
$2$  &  $1.778$   &  $8$  &   $2.998$ \\
$3$  &  $2.035$   &  $9$  &   $3.155$ \\
$4$  &  $2.261$   &  $10$  & $3.304$ \\
$5$  &  $2.466$   &  $11$  & $3.447$ \\
$6$  &  $2.655$   &  $12$  & $3.584$ \\
\end{tabular}
\end{ruledtabular}
\end{table}

If we change $\delta \rightarrow \delta + \Delta \delta$, then the change in the solution $\alpha \rightarrow \alpha_\delta + \Delta \alpha$ of the transcendental equation is approximately given by
\begin{equation}\label{corrtranseq}
\Delta \alpha = - \frac{e^{2 \alpha_\delta^2} \Delta \delta}{4 \alpha_\delta (L_n(4 \alpha_\delta^4) + 2 L_{n-1}^1(4 \alpha_\delta^2))},
\end{equation}
where $\alpha_\delta$ is the solution for $\delta$.

It is noteworthy to mention that we have numerically observed that the last maximum of the function $e^{-x/2}|L_n(x)|$ is monotonically decreasing with $n$. As a consequence, it could be that, for any fixed $\delta$, there exists a value of $n$ such that the last solution to Eq.~\eqref{transeq} could be placed before the last Juddian point. In any case, even in the situation in which this does not hold or when we want $\delta$ above this threshold, there are several strategies to overcome this complication. The most straightforward approach is to consider the analytical extension of the curve fitted for smaller $n$. This actually works since this region is indeed perturbative. A second more complicated approach is introducing $\delta = 1 - F$, with $F$ the fidelity of the perturbative eigenstates given by Eq.~\eqref{DSCsol} in comparison with the exact eigenstates of the QRM. The problem with this approach relies on the lack of an analytically simple expression for the exact eigenstates of the QRM, and thus the fidelity can only be calculated numerically.

\subsection{Non-perturbative ultrastrong/deep strong coupling regime}
\label{sec:npUSCdef}

\begin{figure}[t]
\includegraphics[trim = 8mm 0mm 8mm 7mm, clip, width=0.5\textwidth]{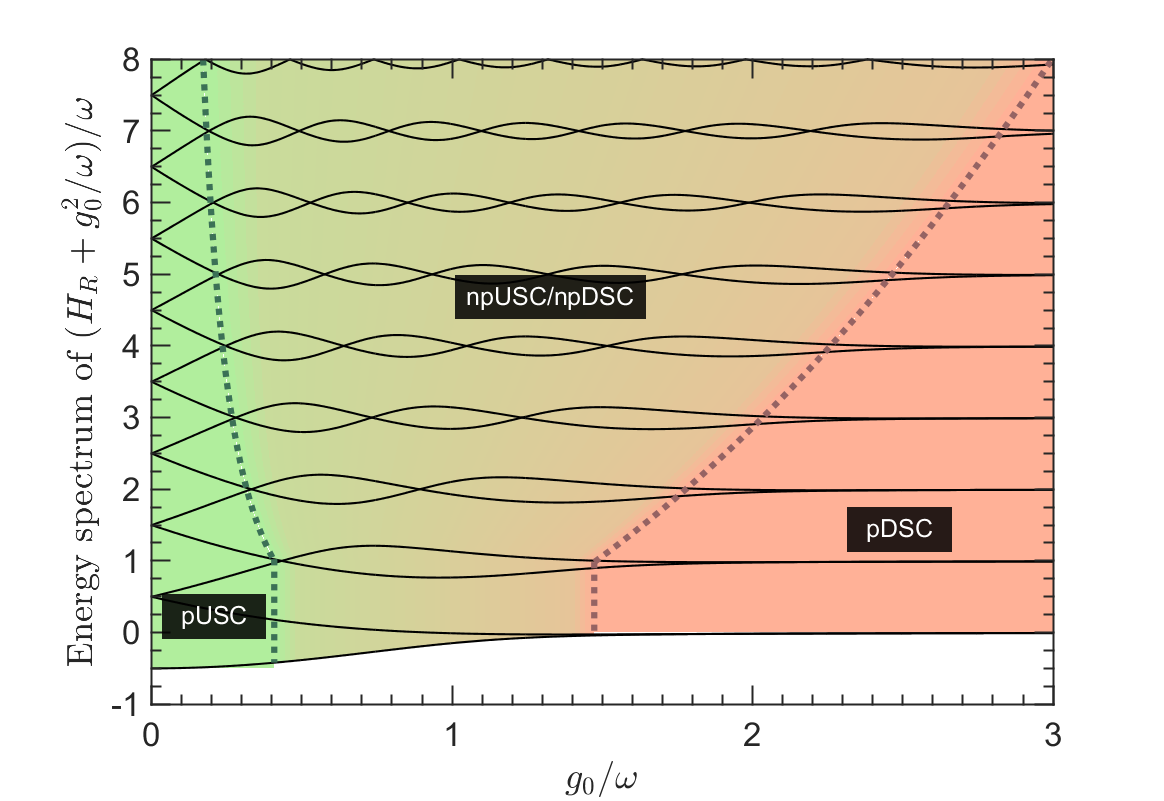}
\caption{Classification of the coupling regimes of the QRM. The region before the first Juddian points stands for the perturbative ultrastrong coupling (pUSC) regime, while the region beyond the last Juddian points represents the perturbative deep strong coupling (pDSC) regime. The intermediate region symbolizes the non-perturbative ultrastrong/deep strong coupling (npUSC/npDSC) regime. The color gradient around the boundaries symbolizes that our spectral classification does not implies an abrupt change in the physical properties of the QRM. Actually, such changes occurs gradually, as shown in Sections~\ref{sec:static} and~\ref{sec:dynam}.}
\label{fig_zones}
\end{figure}

According to the aforementioned results, we are able to classify the coupling regimes of the QRM into three regions, whose boundaries depend not only on the relation between the coupling strength and the natural frequencies of the unperturbed parts, but also on the mean energy that the system can access, as summarized in Fig.~\ref{fig_zones}. The pUSC regime belongs to the region right before the first Juddian points, whose physics are well described by the BS Hamiltonian, which still considers the interaction term as a perturbation. The BS Hamiltonian includes the well-known JC model, i.e., the QRM under the RWA. On the other hand, the pDSC regime belongs to the region beyond the last Juddian points, where there is a role interchange, since the interaction Hamiltonian becomes the main driver of the dynamics, while the bare Hamiltonian is the perturbative term. Finally, between these two coupling regimes, there is a region in which all parts of the Hamiltonian contribute on an equal footing to the dynamics. Then, we can establish this region as the non-perturbative USC (npUSC) regime, or as the non-perturbative DSC (npDSC) regime.

\section{Static properties}
\label{sec:static}

Although our classification seems originally related to the validity of approximate mathematical models, it is associated to physical properties of the QRM which change their behavior qualitatively from region to region. In this section, we will focus on three relevant static properties, namely, the total number of excitations in the system, the photon statistics of the field, and the cavity-qubit entanglement.

\begin{figure}[t]
\centering
\includegraphics[trim = 7mm 7mm 15mm 12mm, clip, width=0.21\textwidth]{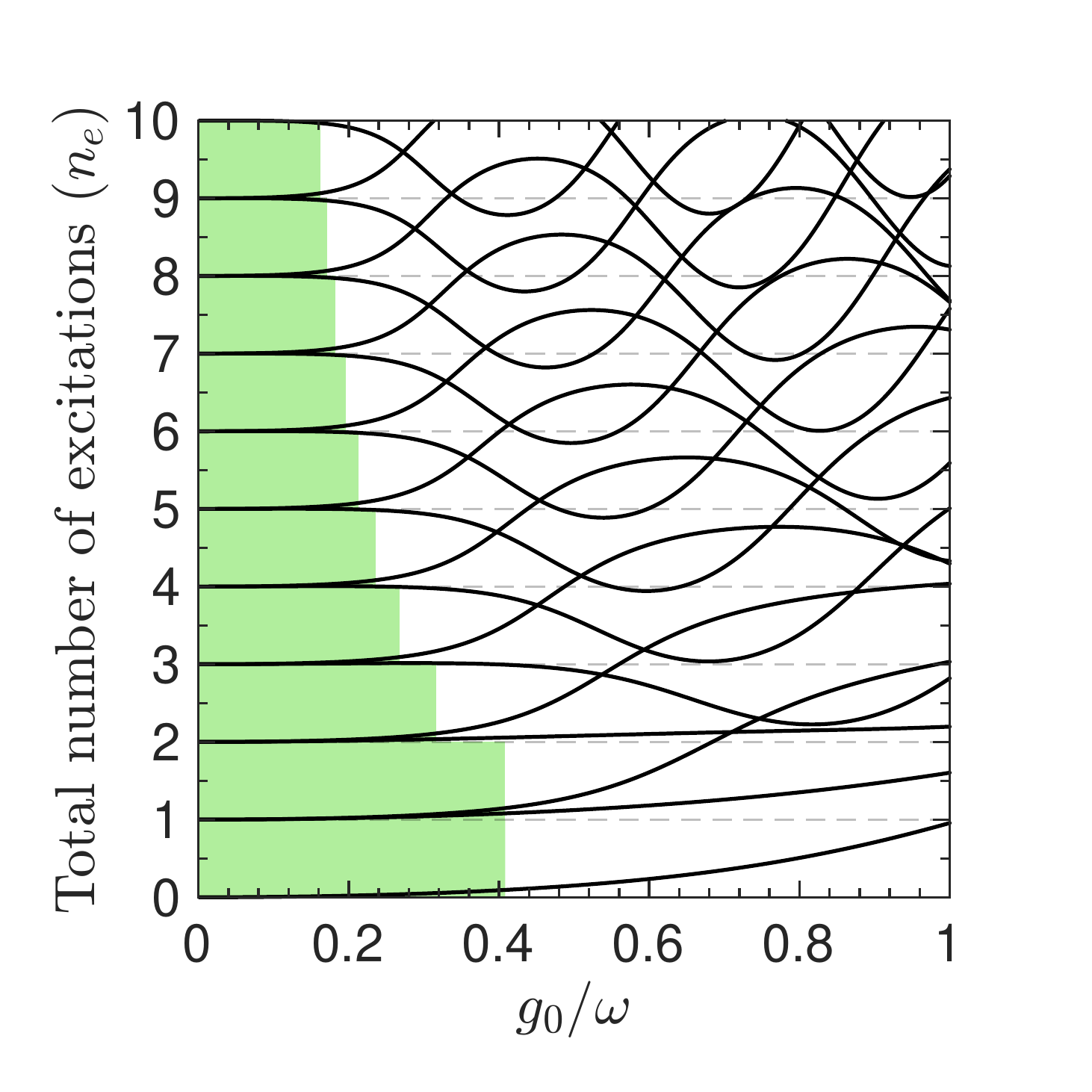}
\includegraphics[trim = 7mm 7mm 15mm 12mm, clip, width=0.21\textwidth]{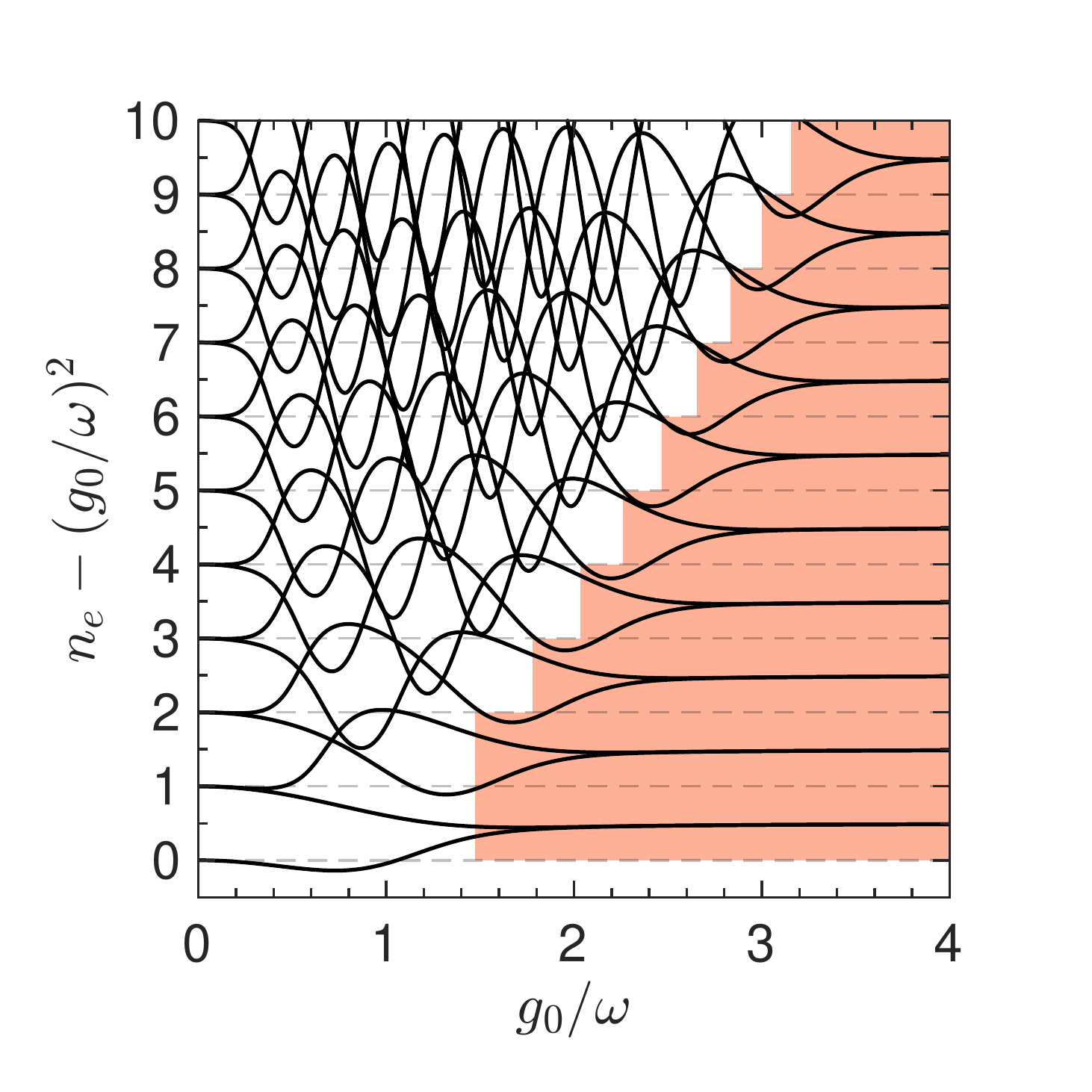}

(a)

\includegraphics[trim = 16mm 20mm 32mm 10mm, clip, width=0.46\textwidth]{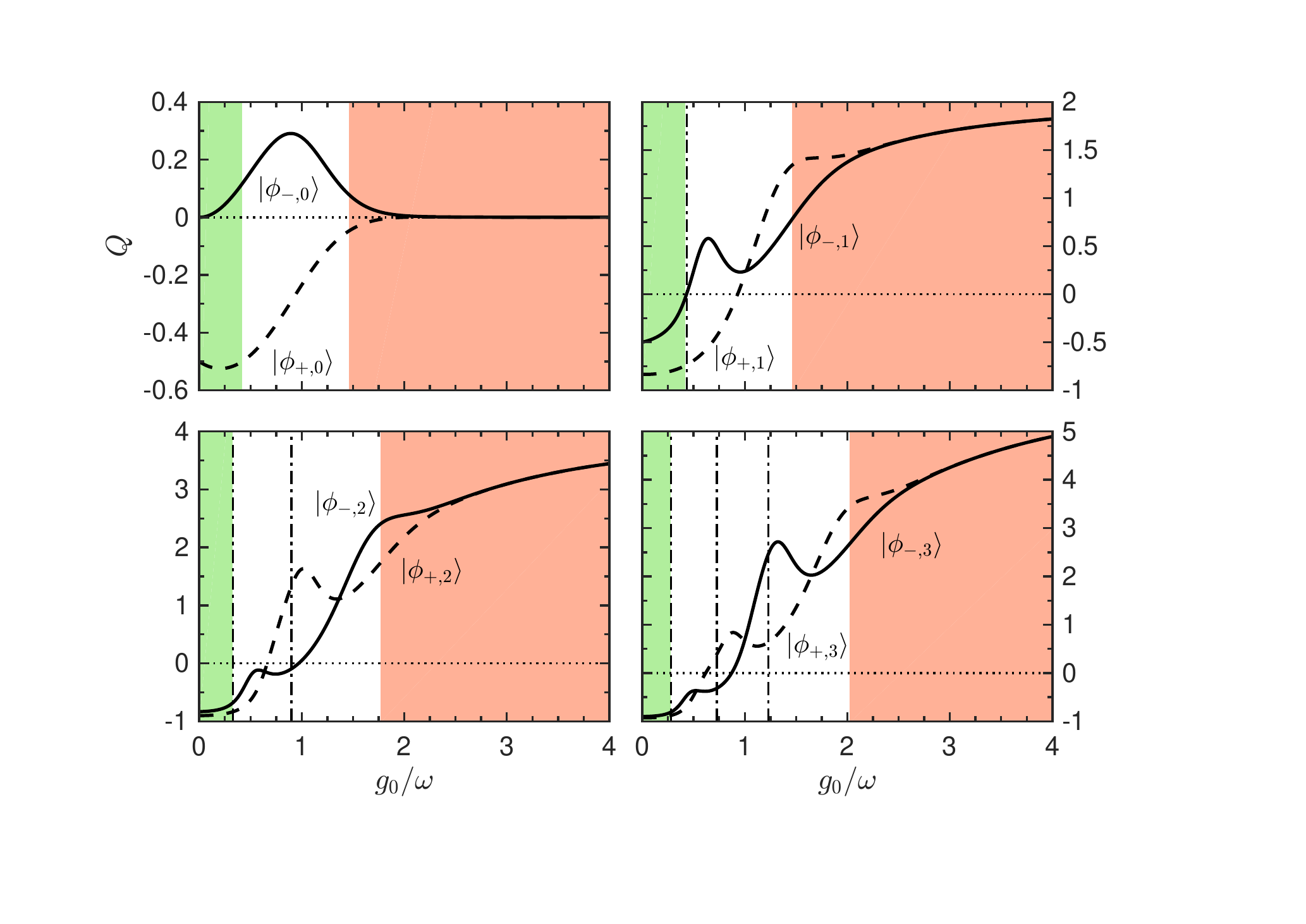}

(b)

\includegraphics[trim = 16mm 20mm 32mm 5mm, clip, width=0.46\textwidth]{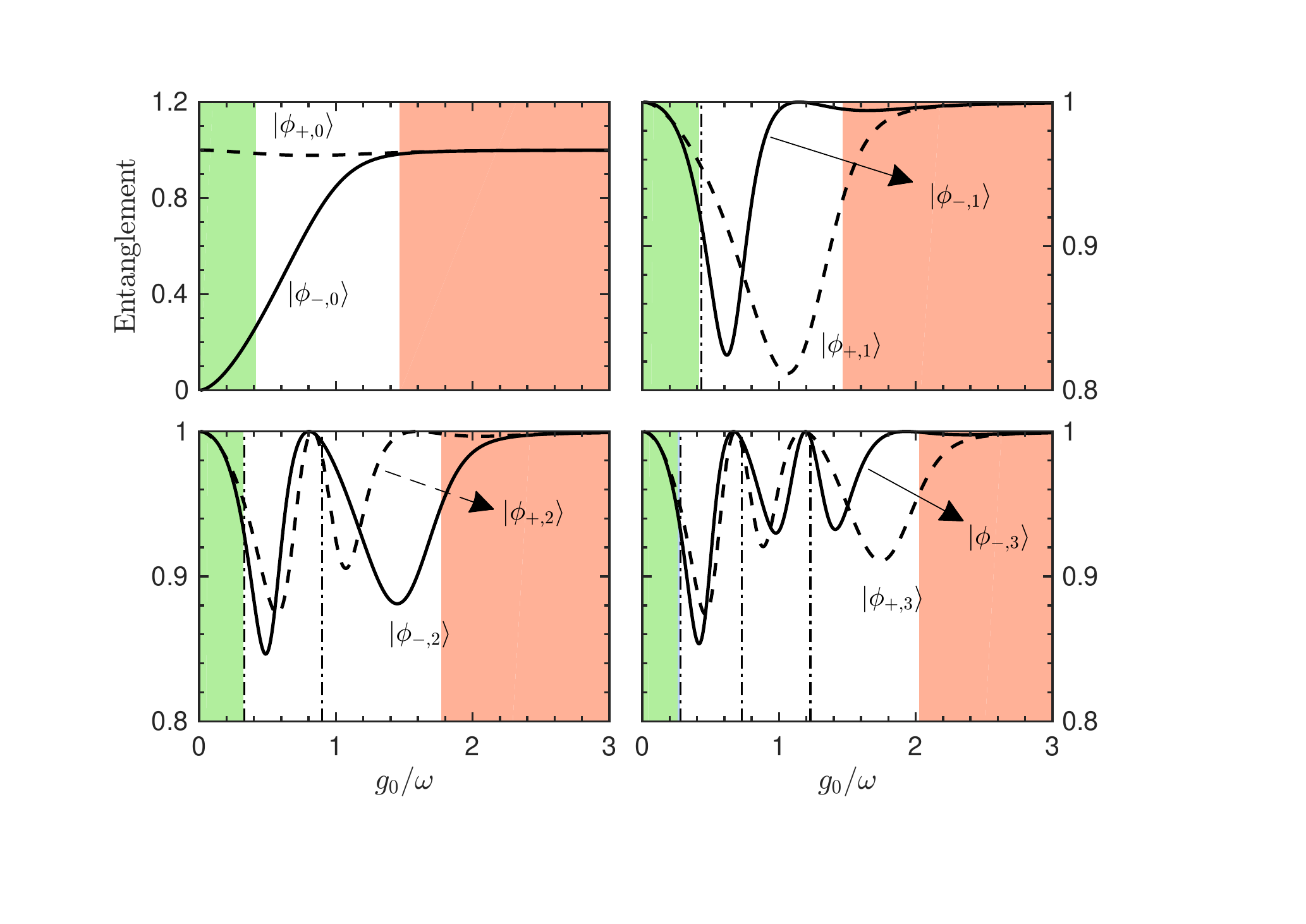}

(c)

\caption{(a) Total number of excitations, (b) Fano-Mandel parameter of the photon distribution of the field, and (c) qubit-field entanglement for the first eigenstates of the QRM in function of $g_{0}/\omega$. The left and right shaded areas stand for the pUSC and pDSC regimes, respectively, while the vertical dashed-dotted lines stand for the energy crossings. For the sake of illustration, we use the same terminology of the pDSC regime to label the eigenstates in (b) and (c).}
\label{neM_figure}
\end{figure}

In Fig.~\ref{neM_figure}(a), we show the total number of excitations ($n_{e} = \langle \hat{n}_e \rangle = \langle a^{\dagger}a + \sigma_{+}\sigma_{-} \rangle$) for each eigenstate of the QRM as function of $g_{0}/\omega$. In the left panel, we observe that the $n_{e}$ remains almost unchanged for the lower-coupling region of the pUSC regime ($g_{0} \ll g_{\text{pUSC}}^{\times} $), which is expected since the BS Hamiltonian which governs the dynamics in this region commutes with $\hat{n}_e$. As we enter in the npUSC/npDSC regime, $n_{e}$ has a non-trivial oscillatory dependency with the coupling strength, which ceases as we approach the pDSC region ($g_{0} \sim g_{\text{pDSC}}^{\times}$), as depicted in the right panel of Fig.~\ref{neM_figure}(a). In the pDSC region, the total number of excitations becomes quasi-degenerate and increases with $(g_{0}/\omega)^{2}$ for the higher-coupling region of this region, with $|n_{e} - (n+1)/2 -  (g_{0}/\omega)^{2}|\leq \delta$ as shown in Appendix~B.

Another physical property with characteristic behavior in each coupling region is the photon statistics of the field. Using the Fano-Mandel parameter $Q$ given by Eq.~\eqref{FM2}  \cite{knight2005}, we distinguish sub-Poissonian ($Q<0$ $-$ genuine nonclassical statistics), Poissonian ($Q=0$), and super-Poissonian ($Q>0$) statistics. Such characteristic behavior is illustrated in Fig.~\ref{neM_figure}(b), where we note that, except for the first two eigenstates that do not have an energy crossing, the field always exhibits a sub-Poissonian and a super-Poissonian photon statistics in the pUSC and pDSC regimes (see Appendix C), respectively, while all kind of photon statistics can be observed in the npUSC/npDSC regime. Moreover, we show in Appendix C that there are transitions in the photon statistics only in the npUSC/npDSC regime.

In Fig.~\ref{neM_figure}(c), we observe that the entanglement between the qubit and the field (von-Neumann entropy of each subsystem \cite{knight2005}) also shows a peculiar behaviour, with the minima only appearing in the npUSC/npDSC regime. In addition, each minimum is always localized between two Juddian points and after the last one. The approximate analytical expressions of the qubit-field entanglement for the pUSC and pDSC regimes are given in Appendix D.

\begin{figure}[t]
\includegraphics[trim = 14mm 21mm 18mm 12mm, clip, width=0.5\textwidth]{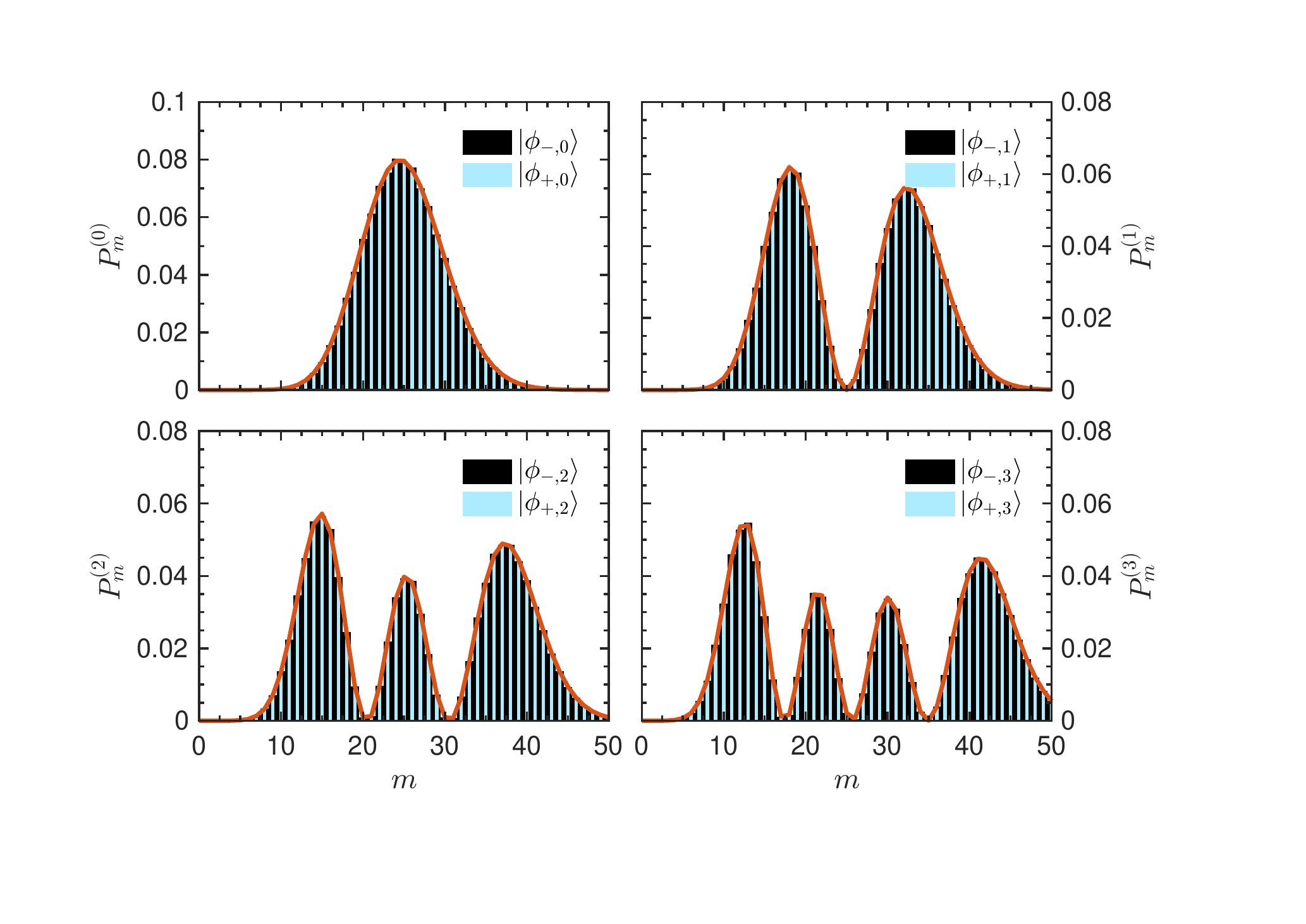}
\caption{Decomposition for $g_{0} \gg g_{\text{pDSC}}^{\times}$ of the field state in the Fock basis, considering $g_{0} = 5\omega$. The bars are computed through the exact eigenstates, while the solid line is obtained by using Eq.~(\ref{DSCsol}).}
\label{decomp}
\end{figure}

Our last remark is related to the decomposition of the field state in the Fock basis \{$\vert m \rangle$\} for the higher-coupling region of the pDSC regime, i.e., for $g_{0} \gg g_{\text{pDSC}}^{\times}$. Using the terminology of the pDSC regime just to label the eigenstates, this decomposition is given by
\begin{equation}
P_{m}^{(\pm,n)} = \text{Tr}(\boldsymbol{1}_{q} \otimes \vert m \rangle \langle m \vert \otimes \vert \phi_{\pm,n} \rangle \langle \phi_{\pm,n} \vert),
\label{dec}
\end{equation}
in which $\boldsymbol{1}_{q} = \sigma_{+}\sigma_{-} + \sigma_{-}\sigma_{+}$ and $\text{Tr}(\bullet)$ is the trace operation. We display $P_{m}^{(\pm,n)}$ in Fig.~\ref{decomp} considering $g_{0} = 5\omega$, in which we firstly recognize that $P_{m}^{(-,n)}$ and $P_{m}^{(+,n)}$ tend toward the same multimodal distribution centered at $m_{c} \simeq (g_{0}/\omega)^{2}$. This can be confirmed by using Eq.~(\ref{DSCsol}), which predicts (red solid line in Fig.~\ref{decomp})
\begin{equation}
P_{m}^{(\pm,n)} = \frac{\alpha^{2\vert m-n \vert}}{e^{\alpha^2}} \frac{\min{(m,n)}!}{\max{(m,n)}!} \left(  L_{\min{(m,n)}}^{\vert m-n\vert} (\alpha^2)   \right)^2,
\label{dec2}
\end{equation}
in which $L_{n}^{a}(x)$ is the generalized Laguerre polynomial and $\alpha = g_{0}/\omega$. Secondly, we can also see that the number of modes of $P_{m}^{(\pm,n)}$, $n+1$, seems to be related to the number of energy crossings between $\vert \phi_{-,n}^{\text{pDSC}} \rangle$ and $\vert \phi_{+,n}^{\text{pDSC}} \rangle$, which is $n$.

It is worth to emphasize that the boundaries of our spectral classification does not imply an abrupt change in the physical properties of the QRM, as noticed in Fig.~\ref{neM_figure}. Actually, such change gradually occurs around the boundaries of the pUSC and pDSC regions, even for dynamical properties as we will see in the next section.

\section{Connection with dynamical properties}
\label{sec:dynam}
As already mentioned in Sec.~\ref{sec:pDSCdef}, the traditional characteristic signature of the DSC regime is not a static property, but a dynamical one. Namely, the appearance of photon-number wave packets that bounce back and forth along a defined parity chain, yielding collapses and revivals of the initial population. In this section, we show how the appearance of this phenomenon is related to our spectral classification.

\begin{figure}[t]
\includegraphics[trim = 10mm 7mm 20mm 13mm, clip, width=0.45\textwidth]{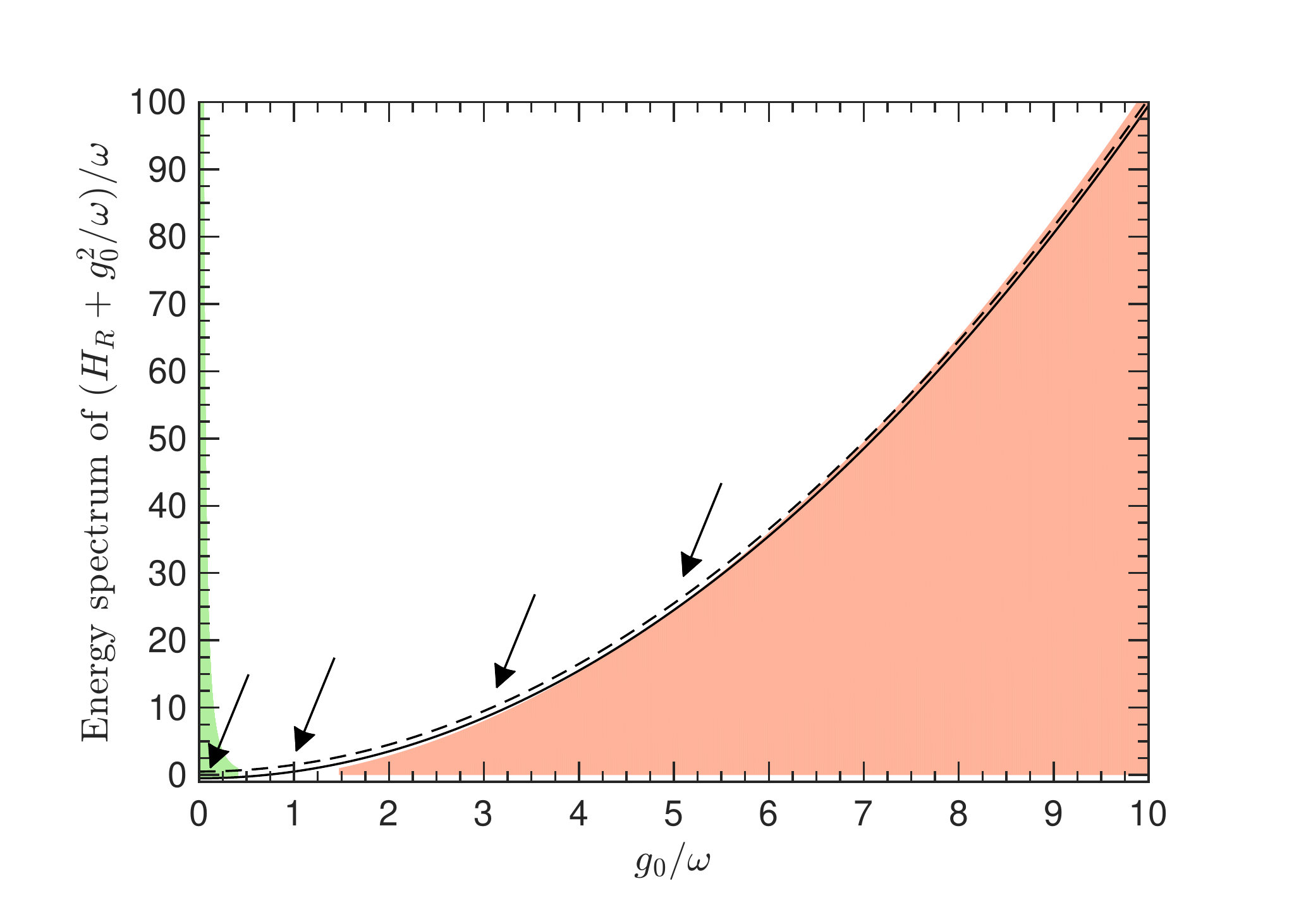}
\includegraphics[trim = 14mm 22mm 20mm 13mm, clip, width=0.45\textwidth]{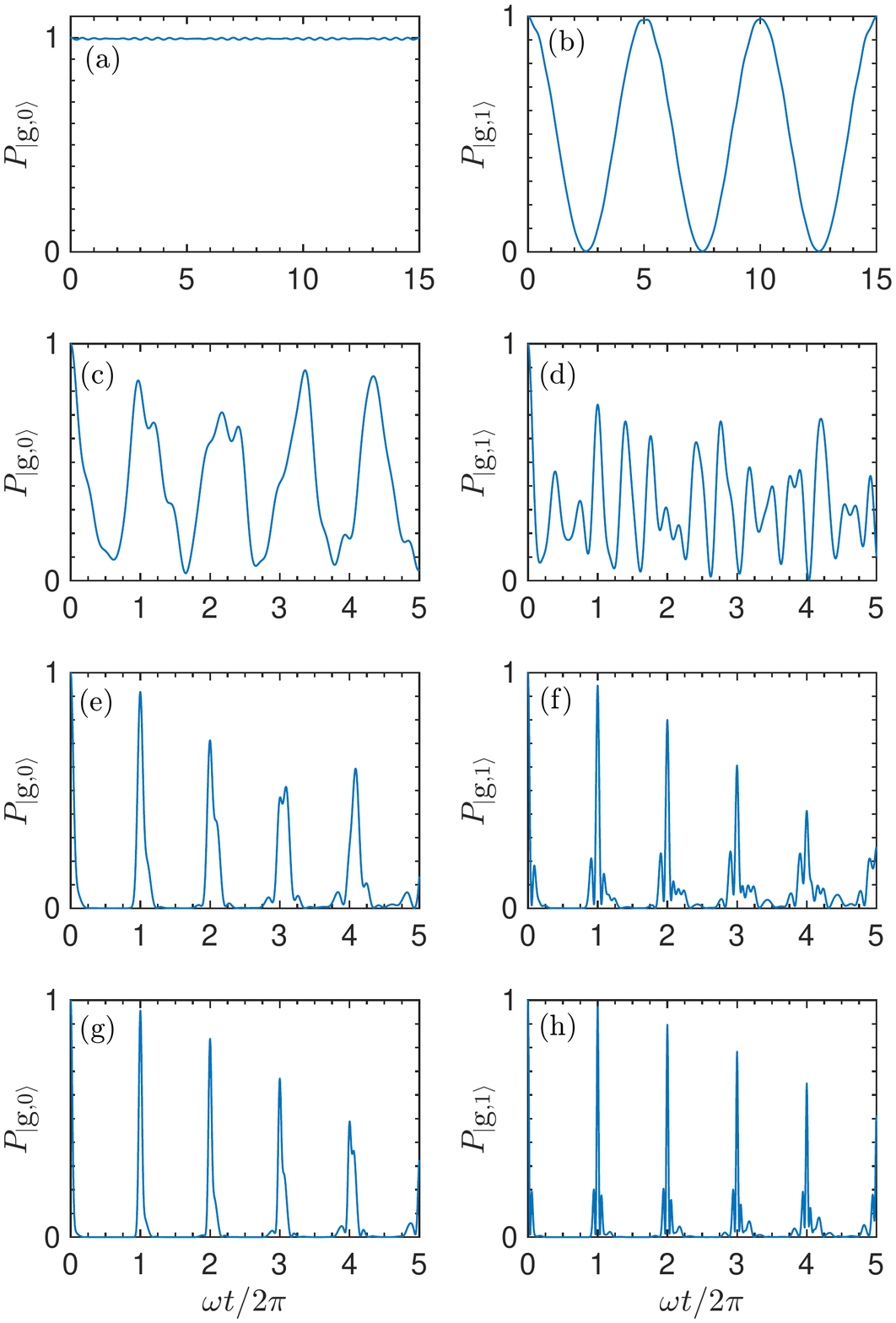}
\caption{(Upper panel) Spectral classification together with the mean energy of the initial states $\vert \text{g},0 \rangle$ (solid line) and $\vert \text{g},1 \rangle$ (dashed line) as function of $g_{0}/\omega$. The arrows indicate the values of $g_{0}$ which are used for the computation of the initial population $P_{\vert \text{g},0\rangle}$ and $P_{\vert \text{g},1\rangle}$, namely, $g_{0}/\omega = 0.1$ in (a) and (b), $g_{0}/\omega = 1$ in (c) and (d), $g_{0}/\omega = 3$ in (e) and (f), and $g_{0}/\omega = 5$ in (g) and (h).}
\label{collrev}
\end{figure}

In the upper panel of Fig.~\ref{collrev}, we show the spectral classification together with the mean energy $\langle \psi _{0} \vert H_{R} \vert \psi _{0} \rangle$ of two initial states, $\vert \text{g},0 \rangle$ (solid line) and $\vert \text{g},1 \rangle$ (dashed line), as function of $g_{0}/\omega$. Considering the values of $g_{0}$ pointed out by the arrows, we computed the initial population $P_{\vert \psi_{0}\rangle}=\langle \psi _{0} \vert e^{-iH_{R}t} \vert \psi _{0} \rangle$. In the pUSC region, $P_{\vert \text{g},0\rangle}$ remains almost constant since $\vert \text{g},0\rangle$ is basically the ground state in that region [Fig.~\ref{collrev}(a)], while $P_{\vert \text{g},1\rangle}$ exhibits Rabi oscillations due to the conservation of the total number of excitation [Fig.~\ref{collrev}(b)]. As we enter in the npUSC/npDSC region, the Rabi oscillations pattern is lost, since the counter-rotating terms introduce a non-trivial oscillatory behavior in the initial population, as shown in Fig.~\ref{collrev}(c)$-$(d). Finally, as we approach the pDSC region [Fig.~\ref{collrev}(e)$-$(f)], the initial population start to present the collapse-revival pattern, which become more prominent as we go inside that region [Fig.~\ref{collrev}(g)$-$(h)].

\section{Conclusion}
\label{sec:conc}

In summary, we have introduced a spectral classification of the coupling regimes of the quantum Rabi model based on the validity of different perturbative approximations, showing that such regimes depend not only on the ratio between the coupling strength and the natural frequencies of the unperturbed parts, but also on the mean energy accesible by the system. Our classification is comprised by three coupling regions, namely the perturbative ultrastrong, the non-perturbative ultrastrong/deep strong and the perturbative deep strong coupling regimes. Remarkably, we have shown that the spectral classification is supported by a clearly divergent behavior of several relevant static physical properties in different coupling regimes. Additionally, we have also tested the suitability of our classification for the usual dynamical properties studied in the literature, which yield the traditional vague USC/DSC division. Therefore, our results clearly answer the long-standing question of providing a founded comprehensible classification of the coupling regimes in the QRM. Moreover, our results also open novel questions which motivate further studies of the mathematical and physical properties of these coupling regions, such as the physical role of the Juddian points in the QRM.

\begin{acknowledgments}
We thank D. Braak, S. Felicetti, G. Romero, J. Casanova, and P. Forn-D\'{i}az for fruitful discussions. This work was supported by the S\~{a}o Paulo Research Foundation (FAPESP) Grants No. 2013/04162-5, 2013/23512-7, and 2014/24576-1, Brazilian National Institute of Science and Technology for Quantum Information (INCT-IQ), CNPq, Spanish MINECO/FEDER Grant FIS2015-69983-P, Basque Government Grant IT986-16, and UPV/EHU UFI 11/55.
\end{acknowledgments}

\appendix

\section{Influence of the higher orders of the BS approximation}
\label{sec:BSexp}
\begin{figure}[t]
\includegraphics[trim = 10mm 7mm 10mm 13mm, clip, width=0.5\textwidth]{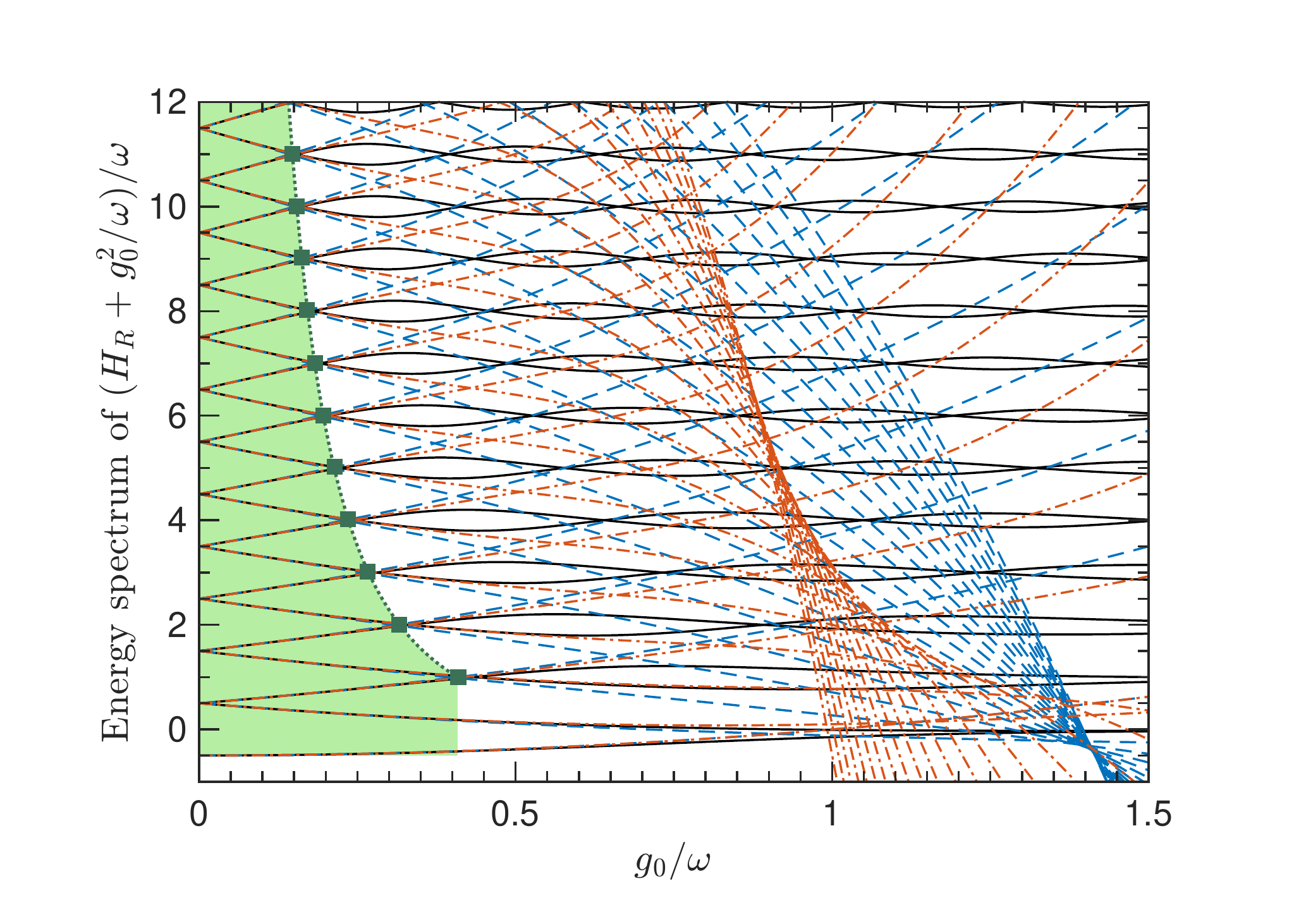}
\caption{Effect of the third-order BS expansion in the pUSC region. We compare the exact eigenenergies of the QRM (solid lines) with the eigenenergies of the second-order BS expansion (dashed lines) and the third-order BS expansion (dashed-dotted lines). We observe that the divergence is still in the first Juddian points, therefore the shaded region still stands for the pUSC regime.}
\label{fig_3order}
\end{figure}

In this Appendix, we discuss the influence of the third order of the BS expansion of the QRM in the definition of the pUSC region. First, let us consider the BS Hamiltonian expanded up to the third order in $g_{0}/\omega$~\cite{forndiaz2016}
\begin{align}\label{3rdorder}
H_{\text{BS}}^{(3)} & = \omega a^{\dagger}a +\frac{\omega}{2}\sigma_z - \omega_{\text{BS}} \left ( \sigma_z a^{\dagger} a  + \frac{1}{2}\right )\nonumber \\ 
& + g(\hat{n}) (a^{\dagger}\sigma^{-}+a \sigma^{+}),
\end{align}
where
\begin{equation}
g(\hat n) = g_{0} \left(1-a^{\dagger}a \frac{\omega_{\text{BS}}}{2\omega} \right)
\end{equation}
is the photon-dependent coupling strength. Notice that the Hamiltonian given by Eq.~\eqref{3rdorder} preserves the number of excitations, i.e., $[H_{\text{BS}}^{(3)}, a^{\dagger}a+\sigma_{+}\sigma_{-}] = 0$. In Fig.~\ref{fig_3order}, the exact eigenenergies of the QRM are depicted and compared with both second and third orders of the BS expansion. One can observe that, even though the third order is more accurate, it still diverges from the correct eigenenergies also after the first Juddian points. Thus, as mentioned in Sec.~\hyperref[sec:pUSCdef]{IIA}, the proposed definition for the pUSC region is not a simple consequence of the second order term, but something deeper related to the breaking of the assumptions for the adiabatic expansion and the point from which the number of excitations starts to be not preserved, as we have seen in the left panel of Fig. \ref{neM_figure}(a).

\section{Total excitations in the perturbative regimes}
\label{sec:totalexc}
The total number of excitations is given by the operator $\hat n_e = a^{\dagger}a + \sigma_{+}\sigma_{-}$. The fact that the number of excitations is preserved in the pUSC regime is a direct consequence of the commutation of this operator with the BS Hamiltonian $[\hat n_e, H_{\text{BS}}^{(2)}]= 0$ (note that this also holds for $H_{\text{BS}}^{(3)}$). 

Let us now compute the mean value of the operator $\hat n_e$ in the pDSC regime, i.e. $\langle \hat{n}_{e}\rangle_{\text{pDSC}} = \langle \phi_{\pm,n}^{\text{pDSC}}|\hat n_e | \phi_{\pm,n}^{\text{pDSC}} \rangle$. Hence,
\begin{align}\label{nepdsc}
\langle \hat{n}_{e}&\rangle_{\text{pDSC}} =  \frac{1}{2} \left[\langle n | \mathcal{D}^{\dagger}(\alpha)a^{\dagger} a \mathcal{D}(\alpha)|n\rangle \pm  \langle n | \mathcal{D}^2(\alpha)|n\rangle +1 \right. \nonumber \\ 
& + \left. \langle n | \mathcal{D}^{\dagger}(-\alpha)a^{\dagger} a \mathcal{D}(-\alpha)|n\rangle  \pm  \langle n | \mathcal{D}^2(-\alpha)|n\rangle \right],
\end{align}
with $\alpha = g_{0}/\omega$. In order to compute the first term, we use that $\langle n | \mathcal{D}^{\dagger}(x)a^{\dagger} a \mathcal{D}(x)|n\rangle = \langle n | (a^{\dagger}-x)(a - x) |n\rangle = n +x^2$. The second term is given by Eq.~\eqref{D2}, so that $\langle n | \mathcal{D}^2(x)|n\rangle=e^{-2x^2} {}_1F_1 (-n,1;4x^2)$, in which ${}_1F_1 (a,b;z)$ is the Kummer's confluent hypergeometric function. By using that these functions are real valued, we obtain that
\begin{align} \label{nepdsc2}
\langle \hat{n}_{e}\rangle_{\text{pDSC}} - \alpha^2  & = n + \frac{1}{2} \pm e^{-2\alpha^2} {}_1F_1 (-n,1;4\alpha^2) \nonumber \\
&= n + \frac{1}{2} \pm \underbrace{e^{-2\alpha^2} L_{n}(4\alpha^2)}_{\leq \delta},
\end{align}
where we have made use of Kummer's transformation $L^{(\alpha)}_n(z) = \binom{n+\alpha}{n} {}_1F_1 (-n,\alpha + 1;z)$~\cite{nist}. Therefore, effectively, the variation in $\langle \hat{n}_{e}\rangle_{\text{pDSC}} - \alpha^2$ is exponentially suppressed when $\alpha \rightarrow \infty$, and upper-bounded by $\delta$ in the pDSC region.

\section{Fano-Mandel parameter in pUSC and pDSC}
\label{sec:fano}
In this Appendix, we compute the photon statistics of the QRM eigenstates in the pUSC and pDSC regimes through the Fano-Mandel parameter
\begin{equation} \label{FM2}
Q = \frac{\langle\hat{n}^2 \rangle-\langle \hat{n} \rangle^2}{\langle \hat{n}  \rangle} - 1.
\end{equation} 
Let us recall that the photon distribution is classified as sub-Poissonian ($Q<0$  $-$ genuine nonclassical statistics), Poissonian ($Q=0$), and super-Poissonian ($Q>0$).

\subsection{Perturbative USC regime}
\label{sec:fanoUSC}
First, we must compute the photon distributions of the eigenstates $|\phi^{\text{BS}}_{n,\pm}\rangle$, which is defined by $P_m(\phi^{\text{BS}}_{n,\pm}) = |\langle \text{g}, m|\phi^{\text{BS}}_{n,\pm}\rangle|^2+|\langle \text{e}, m|\phi^{\text{BS}}_{n,\pm}\rangle|^2$. In order to perform the calculation, it is useful noticing that $\mathcal{U}^{\dagger} |\text{g}, m\rangle = \frac{1}{\sqrt{m!}} (\mathcal{U}^{\dagger} a^{\dagger \,m} \mathcal{U}) \, \mathcal{U}^{\dagger} | \text{g}, 0\rangle$, with $\mathcal{U}$ given by Eq.~\eqref{unitrans}. By using the Baker-Campbell-Hausdorff formula to second order,
\begin{align}\label{part1}
\mathcal{U}^{\dagger} a^m\mathcal{U} &  = a^m + [a^m, \mathcal{H}(\alpha)] \nonumber \\
&+ \frac{1}{2}\left[[a^m,\mathcal{H}(\alpha)],\mathcal{H}(\alpha)\right] + \mathcal{O}(\alpha^3),
\end{align}
with,
\begin{equation} \label{halpha}
\mathcal{H}(\alpha)= (\alpha/2)(a \sigma_{-} - a^{\dagger} \sigma{+}) + (\alpha^{2}/4)(a^{2} - a^{\dagger 2})\sigma_{z},
\end{equation}
and $\alpha = g_{0}/\omega$.

It is straightforward to prove the useful expressions $[a^{\dagger}, a^m]=-m a^{m-1}$ and $[a^{\dagger \, 2}, a^m]=-m (a^{\dagger} a^{m-1} + a^{m-1} a^{\dagger})$, which may be used to compute, to the second order, the commutator
\begin{align}\label{comm1}
[a^m, \mathcal{H}(\alpha)&]  = -\frac{\alpha}{2}m a^{m-1} \sigma^{\dagger} \nonumber \\
& -\frac{\alpha^2}{4}m \sigma_z (a^{\dagger} a^{m-1} + a^{m-1} a^{\dagger}) + \mathcal{O}(\alpha^3).
\end{align} 
Let us now compute the second commutator of Eq.~\eqref{part1} to the second order 
\begin{equation}\label{comm2}
\left[[a^m,\mathcal{H}(\alpha)],\mathcal{H}(\alpha)\right] = -\frac{\alpha^2}{4}m a^m \sigma_z + \mathcal{O}(\alpha^3).
\end{equation}
Therefore, by replacing Eqs.~\eqref{comm1} and \eqref{comm2} into Eq.~\eqref{part1}, we obtain
\begin{align}\label{comm22}
\mathcal{U}^{\dagger} a^m\mathcal{U}&= a^m - \frac{\alpha}{2}m a^{m-1} \sigma^{\dagger} - \frac{\alpha^2}{4}m a^m \sigma_z \nonumber \\
 &- \frac{\alpha^2}{4} m \sigma_z (a^{\dagger} a^{m-1} + a^{m-1} a^{\dagger})  + \mathcal{O}(\alpha^3).
\end{align} 

Now, we have to compute $\mathcal{U}^{\dagger}  |\text{g}, 0\rangle$, also to the second order in $\alpha$, i.e., $\mathcal{U}^{\dagger} = \mathbbm{1} - \mathcal{H}(\alpha)+\frac{1}{2}\mathcal{H}(\alpha)^2 + \mathcal{O}(\alpha^3)$, which yields after normalization
\begin{align}\label{part2}
\mathcal{U}^{\dagger}  |\text{g}, 0\rangle &= \left( 1- \frac{\alpha^2}{8} \right)|\text{g},0\rangle + \frac{\alpha}{2} |\text{e},1\rangle \nonumber \\
&-\frac{\alpha^2 \sqrt{2}}{4} |\text{g},2\rangle  + \mathcal{O}(\alpha^3).
\end{align}
By using this together with Eq.~\eqref{part1}, we obtain
\begin{align}
\mathcal{U}^{\dagger} |\text{g}, m\rangle & = \left [1-(m+1)\frac{\alpha^2}{8} \right] |\text{g},m\rangle + \frac{\alpha}{2}\sqrt{m+1} |\text{e},m+1\rangle \nonumber \\
&-\frac{\alpha^2}{4}\left( \sqrt{(m+1)(m+2)} |\text{g}, m+2\rangle \right. \nonumber \\
& \left.- \sqrt{m(m-1)} |\text{g},m-2\rangle  \right).
\end{align}
Analogously, 
\begin{align}
\mathcal{U}^{\dagger} |\text{e}, m\rangle &= \left (1+m\frac{\alpha^2}{4} \right) |\text{e},m\rangle - \frac{\alpha}{2}\sqrt{m} |\text{g},m-1\rangle \nonumber \\
&+ \frac{\alpha^2}{4}\left( \sqrt{(m+1)(m+2)} |\text{e}, m+2\rangle \right. \nonumber \\
&- \left. \sqrt{m(m-1)} |\text{e},m-2\rangle  \right).
\end{align}
The scalar products of these states with respect to the state given by Eq.~\eqref{pnBS} yields
\begin{align}
\langle \text{g}, m | \mathcal{U} | +, n\rangle & = \left ( 1 + m \frac{\alpha^2}{4} \right ) \cos \left(\frac{\theta_{m+1}}{2}\right)\delta_{n,m+1}\nonumber\\
& - \frac{\alpha}{2}\sqrt{m} \sin \left(\frac{\theta_{m-1}}{2}\right) \delta_{n,m-1} \nonumber\\ & + \frac{\alpha^2}{4} \left [\sqrt{(m+1)(m+2)} \cos\left(\frac{\theta_{m+3}}{2}\right)\delta_{n,m+3} \right. \nonumber \\
& \left.- \sqrt{m(m-1)} \cos\left(\frac{\theta_{m-3}}{2}\right) \delta_{n,m-3} \right].  \label{pos1} \\
\langle \text{e}, m | \mathcal{U} | +, n\rangle & = \left [ 1 - (m+1) \frac{\alpha^2}{8} \right ] \sin \left(\frac{\theta_{m}}{2}\right)\delta_{n,m} \nonumber \\
&+ \frac{\alpha}{2}\sqrt{m+1} \cos \left(\frac{\theta_{m+2}}{2}\right) \delta_{n,m+2} \nonumber\\ & - \frac{\alpha^2}{4} \left [\sqrt{(m+1)(m+2)} \cos\left(\frac{\theta_{m+2}}{2}\right)\delta_{n,m+2} \right. \nonumber \\
& \left. - \sqrt{m(m-1)} \cos\left(\frac{\theta_{m-2}}{2}\right) \delta_{n,m-2} \right ]. \label{pos2}
\end{align}
The sum of the squares of these elements gives the photon distributions $P_m(\phi^{\text{BS}}_{n,+})$, for which we need to use the expressions $\sin^2\frac{\theta_n}{2} = \frac{1}{2}(1-\frac{\alpha\sqrt n}{2})$ and $\cos^2 \frac{\theta_n}{2} = \frac{1}{2}(1+\frac{\alpha\sqrt n}{2})$. 

Now, we want to compute the first and second moments of the distribution
\begin{align}
\langle \hat{n}\rangle & = \sum_{m=0}^{\infty} m P^{+,n}_m = (n-\frac{1}{2}) - \frac{\alpha \sqrt{n}}{5} \nonumber \\
&+\frac{1}{8}(5-7n+2n^2) + \mathcal{O}(\alpha^3), \\
\langle \hat{n}^2\rangle & = \sum_{m=0}^{\infty} m^2 P^{+,n}_m= (n^2-n+\frac{1}{2}) +\frac{\alpha}{4}(1-2n)\sqrt{n}\nonumber \\
&+\frac{\alpha^2}{8}(2n^3-10 n^2+17 n -5) + \mathcal{O}(\alpha^3).
\end{align}
In order to prove that the distribution is sub-Poissonian, it is sufficient to study the sign of 
\begin{align}
\langle \hat{n}^2\rangle-\langle \hat{n}\rangle^2-\langle \hat{n}\rangle &= (\frac{3}{4}-n) + \frac{\alpha \sqrt{n}}{4} \nonumber \\
&-\frac{\alpha^2}{16}(4n^3-8n^2-13n+10),
\end{align}
which can be straightforwardly proven to be negative in the pUSC regime, i.e., assuming that $0\leq \alpha \leq 1/\sqrt{2(2n+1)}$. The cubic polynomial is negative when $n = 1, 2$ and positive when $n \geq 3$. The first case can be directly checked. In the second case, $\langle \hat{n}^2\rangle-\langle \hat{n}\rangle^2-\langle \hat{n}\rangle \leq  (\frac{3}{4}-n) + \sqrt{\frac{n}{8(2n+1)}}\leq (\frac{3}{4}-n) + \frac{1}{4} < 0$, which finally proves that the photon distribution of the states $|\phi^{\text{BS}}_{n,+}\rangle$ is sub-Poissonian. In order to extend it to the states $|\phi^{\text{BS}}_{n,-}\rangle$, it is only necessary to apply the substitutions $\sin\frac{\theta_n}{2} \rightarrow -\cos \frac{\theta_n}{2}$ and $\cos \frac{\theta_n}{2}\rightarrow\sin\frac{\theta_n}{2}$ in Eqs.~\eqref{pos1} and \eqref{pos2} and proceed analogously. This yields
\begin{align}
\langle \hat{n}^2\rangle-\langle \hat{n}\rangle^2-\langle \hat{n}\rangle &= (\frac{3}{4}-n) - \frac{\alpha \sqrt{n}}{4} \nonumber \\
&-\frac{\alpha^2}{16}(4n^3-8n^2-13n+10),
\end{align}
which is also negative for $n\geq 1$. This concludes the proof.

\subsection{Perturbative DSC regime}
\label{sec:fanoDSC}
Here, we prove that the photon distribution of the eigenstates of the QRM in the DSC regime is super-Poissonian. To achieve it, we proceed similarly to the previous subsection, assuming that in DSC the eigenstates are correctly described by Eq.~\eqref{DSCsol}. It is straightforward to see that
\begin{align}
\langle \text{g}, m \vert \phi^{\text{pDSC}}_{\pm,n} \rangle & = \frac{1}{2} (\langle m | \mathcal{D}(-\alpha)|n\rangle \mp \langle m | \mathcal{D}(\alpha)|n\rangle),\\
\langle \text{e}, m \vert \phi^{\text{pDSC}}_{\pm,n} \rangle & = \frac{1}{2} (\langle m | \mathcal{D}(-\alpha)|n\rangle \pm \langle m | \mathcal{D}(\alpha)|n\rangle),
\end{align}
where
\begin{align}\label{cohst}
\langle m | \mathcal{D}(\alpha)|n\rangle = \sqrt{\frac{m!}{n!}}e^{-\frac{1}{2}\alpha^2} \alpha^{m+n} \nonumber \\
\times\sum_{k=0}^{\min (n,m)}\frac{(-1)^{n-k}}{(m-k)!} \binom{n}{k}\alpha^{-2k}
\end{align}

Let us start by computing the photon distribution for $\vert \phi^{\text{pDSC}}_{+,n} \rangle$, which means that $P_m(\phi^{\text{pDSC}}_{n,+}) = |\langle \text{g}, m|\phi^{\text{pDSC}}_{n,+}\rangle|^2+|\langle \text{e}, m|\phi^{\text{pDSC}}_{n,+}\rangle|^2$ is given by
\begingroup\makeatletter\def\f@size{9}\check@mathfonts
\def\maketag@@@#1{\hbox{\m@th\large\normalfont#1}}%
\begin{eqnarray*}
P_m(\phi^{\text{pDSC}}_{n,+})  = \left |\sqrt{\frac{m!}{n!}}\frac{e^{-\frac{1}{2}\alpha^2} \alpha^{m+n}}{2}\sum_{k=0}^{n}\frac{(-1)^{n-k}}{(m-k)!} \binom{n}{k}\alpha^{-2k}\right |^2 \nonumber\\
\times  \underbrace{((1-(-1)^{m+n})^2+(1+(-1)^{m+n})^2)}_{=4}= |\langle m | \mathcal{D}(\alpha)|n\rangle)|^2
\end{eqnarray*}
\endgroup
Taking this into account, the computation of the first and second moments is straightforward, since
\begin{align}
\langle \hat{n} \rangle &= \sum_{m=0}^{\infty} m P^{+,n}_m = \sum_{m=0}^{\infty} \langle n|\mathcal{D}^{\dagger}(\alpha) \underbrace{|m\rangle m \langle m|}_{\hat{n} = a^{\dagger}a} \mathcal{D}(\alpha)|n\rangle \nonumber\\
& = n + \alpha^2, \\
\langle \hat{n}^2 \rangle &= \sum_{m=0}^{\infty} m^2 P^{+,n}_m = \sum_{m=0}^{\infty} \langle n|\mathcal{D}^{\dagger}(\alpha) \underbrace{|m\rangle m^2 \langle m|}_{\hat{n}^2 = (a^{\dagger}a)^2 } \mathcal{D}(\alpha)|n\rangle \nonumber \\
&= n^2 + \alpha^4 + \alpha^2 (4n+1)
\end{align}
Therefore, in order to prove that the distribution is super-Poissonian for $\alpha \gtrsim g_{\text{pDSC}}^{\times}/\omega $, we have to study the sign of
\begin{equation} \label{FMpDSC}
\langle \hat{n}^2\rangle-\langle \hat{n}\rangle^2-\langle \hat{n}\rangle = n(2 \alpha^2 - 1) > 0,
\end{equation}
which proves it, since $\alpha$ is always bigger than $1/\sqrt{2}$ in the pDSC regime (see Table~\ref{pDSCpoints}). For the case of the eigenstates $\vert \phi^{\text{pDSC}}_{-,n} \rangle$, we only need to notice that the photon distribution is exactly the same, hence Eqs.~\eqref{cohst}-\eqref{FMpDSC} also hold, which concludes the proof.

\section{Cavity-qubit entanglement}
\label{sec:entang}

In this Appendix, we compute the cavity-qubit entanglement via the von-Neumann entropy in the pUSC and pDSC regimes, which allows us to analytically prove the numerical observations in Sec.~\hyperref[sec:static]{III}.

\subsection{Von-Neumann entropy in the pUSC regime} 
\label{sec:entUSC}
We have to compute the reduce density matrix for the qubit system. By using the Baker-Haussdorff-Campbell formula,
\begin{align} \label{contbh}
\mathcal{U} |\pm, &n\rangle\langle \pm n| \mathcal{U}^{\dagger}  = |\pm, n\rangle\langle \pm n| + [\mathcal{H}(\alpha),|\pm, n\rangle\langle \pm n|]\nonumber \\
&+\frac{1}{2}[\mathcal{H}(\alpha),[\mathcal{H}(\alpha),|\pm, n\rangle\langle \pm n|]] + \mathcal{O}(\mathcal{H}(\alpha)^3),
\end{align}
in which $\alpha = g_{0}/\omega$ with $\mathcal{H}(\alpha)$ given by Eq.~\eqref{halpha}. The reduced density matrix is obtained by tracing out the bosonic degrees of freedom, i.e., $\rho_{n,\pm} = \text{Tr}_{\text{cav}} (\mathcal{U} |\pm, n\rangle\langle \pm n| \mathcal{U}^{\dagger} )$. As usual, let us first consider the states $| +, n\rangle$. Then, the contribution to the reduced density matrix due to the first term in Eq.~\eqref{contbh} is 
\begin{align}
\rho^{(1)} &=  \text{Tr}_{\text{cav}} (|+, n\rangle\langle + n| )  \nonumber \\
&= \cos^2 \frac{\theta_n}{2} |e\rangle\langle e| + \sin^2 \frac{\theta_n}{2} |g\rangle\langle g| .
\end{align}
For the second term, it is straightforward to prove that $\rho^{(2)} = \text{Tr}_{\text{cav}} ([\mathcal{H}(\alpha),|+, n\rangle\langle + n|] ) = \mathcal{O}(\alpha^3)$, so it will not be considered. 

Finally, for the third term we must only consider the influence of the Hamiltonian term $\frac{\alpha}{2} (a \sigma^{-}-a^{\dagger}\sigma^{+})$, since we are working in $\mathcal{O}(\alpha^2)$. Let us notice that the double commutator can be rewritten as $[\mathcal{H}(\alpha),[\mathcal{H}(\alpha),|\pm, n\rangle\langle \pm n|]] = \{ |\pm, n\rangle\langle \pm n|,\mathcal{H}(\alpha)^2 \} - 2\mathcal{H}(\alpha)|\pm, n\rangle\langle \pm n|\mathcal{H}(\alpha) $, so let us compute both terms separately. The anti-commutator yields
\begin{align}
\text{Tr}_{\text{cav}} ( \{ |+, n\rangle\langle +, n|,\mathcal{H}&(\alpha)^2 \}) = -\frac{\alpha^2}{4}\left[(n-1)\cos^2\frac{\theta_n}{2}  |\text{e}\rangle\langle \text{e}| \right. \nonumber \\
&\left.+(n+1)\sin^2\frac{\theta_n}{2}  |\text{g}\rangle\langle \text{g}|\right].
\end{align}
Analogously, one obtains 
\begin{align}
\text{Tr}_{\text{cav}} (\mathcal{H}(\alpha) |+, n\rangle\langle +, n| \mathcal{H}&(\alpha)) = -\frac{\alpha^2}{4}\left[(n+1)\sin^2\frac{\theta_n}{2}  |\text{e}\rangle\langle \text{e}| \right.\nonumber \\
&\left.+(n-1)\cos^2\frac{\theta_n}{2}  |\text{g}\rangle\langle \text{g}|\right],
\end{align} 
such that the total contribution to the $\rho^{(3)}$, using that $\sin^2\frac{\theta_n}{2} =\cos^2\frac{\theta_n}{2} =\frac{1}{2} + \mathcal{O}(\alpha)$, is given by
\begin{equation}
\rho^{(3)} =  \frac{\alpha^2}{2}  (|\text{e}\rangle\langle \text{e}|- |\text{g}\rangle\langle \text{g}|) = \frac{\alpha^2}{2} \sigma_z. 
\end{equation}
Therefore, the total density matrices for $n>0$ are given by
\begin{align}
\rho_{n,\pm}  & = \left(\frac{1}{2} \pm \frac{\alpha \sqrt{n}}{4} +\frac{\alpha^2}{4} \right) |\text{e}\rangle\langle \text{e}| \nonumber \\
&+ \left(\frac{1}{2} \mp \frac{\alpha \sqrt{n}}{4} -\frac{\alpha^2}{4} \right) |\text{g}\rangle\langle \text{g}|. 
\end{align}

We can see that the entanglement is maximum for $\alpha \approx 0$, as numerically observed. The von-Neumann entropy $S(\rho) = - \sum_k \lambda_k \log_{2} \lambda_k$, where $\lambda_{k}$ are the eigenvalues of $\rho$, is
\begin{equation}
S(\rho_{n,\pm}) = 1 - \frac{n \alpha^2}{8} + \mathcal{O}(\alpha^3).
\end{equation}

\subsection{Von-Neumann entropy in the pDSC regime} 
\label{sec:entDSC}
Let us take the states describing the the cavity-qubit system in the pDSC regime given by Eq.~\eqref{DSCsol}, and trace out the bosonic degrees of freedom
\begin{align}
\text{Tr}_{\text{cav}}\left( |\phi_{\pm,n}^{\text{pDSC}}\rangle\langle \phi_{\pm,n}^{\text{pDSC}}|\right) &= \frac{1}{2} \left [|+\rangle\langle +| + |-\rangle\langle -| \right. \nonumber \\
&\pm |+\rangle\langle -| \langle n | \mathcal{D}^2(-\alpha)|n\rangle \nonumber \\
& \left. \pm  |-\rangle\langle +| \langle n | \mathcal{D}^2(\alpha)|n\rangle \right].
\end{align}
Therefore, the aim here is to compute $\langle n | \mathcal{D}^2(\alpha) |n\rangle = \langle n | \mathcal{D}(2\alpha) |n\rangle = \langle n | \mathcal{D}(2\alpha) \frac{(a^{\dagger})^n}{\sqrt{n!}} \mathcal{D}^{\dagger}(2\alpha) \mathcal{D}(2\alpha) |0\rangle$. By using that $\mathcal{D}(2\alpha) a^{\dagger} \mathcal{D}^{\dagger}(2\alpha) = a^{\dagger}-2\alpha$ and the Newton's binomial theorem,
\begin{align}
\langle n | \mathcal{D}^2(\alpha) |n\rangle &= \frac{1}{\sqrt{n!}} \langle n | \sum_{k=0}^n \binom{n}{k} (-2\alpha)^{n-k} a^{\dagger \, k} |2\alpha \rangle \nonumber \\
& = \sum_{k=0}^n \binom{n}{k}(-2\alpha)^{n-k} \frac{\langle n-k |2\alpha\rangle}{\sqrt{(n-k)!}},
\end{align}
where we have used that $a^k|n\rangle = \sqrt{\frac{n!}{(n-k)!}}|n-k\rangle$. Then, employing $\langle n-k | 2\alpha\rangle = e^{-2\alpha^2}\frac{(2\alpha)^{n-k}}{\sqrt{(n-k)!}}$, we have that
\begin{align}\label{D2}
\langle n | \mathcal{D}^2(\alpha) |n\rangle&= e^{-2\alpha^2} \sum_{k=0}^n \binom{n}{k} (-1)^k \frac{(2\alpha)^{2k}}{k!} \nonumber \\
&= e^{-2\alpha^2} {}_1F_1 (-n,1;4\alpha^2),
\end{align}
where ${}_1F_1(a,b;z)$ is the Kummer confluent hypergeometric function~\cite{nist}. 

Therefore, the two eigenvalues of the reduced density matrix for the qubit are
\begin{equation}
\lambda_{\pm}(\phi_{\pm,n}^{\text{pDSC}}) = \frac{1}{2} \left[1 \pm e^{-2\alpha^2} {}_1F_1(-n,1;4\alpha^2)\right].
\end{equation}
Notice that the eigenvalues only depend on the quantum number $n$. Finally, we can compute the entropy $S=-\frac{1}{2}(1-y) \log_{2} (\frac{1-y}{2})-\frac{1}{2}(1+y) \log_{2} (\frac{1+y}{2}) = 1 - \frac{y^2}{2} + \mathcal{O}(y^4)$, with $y = e^{-2\alpha^2}{}_1F_1(-n,1;4\alpha^2)\ll1$ in the pDSC region. Hence, the entropy is given by
\begin{equation}
S = 1 - \frac{1}{2} \underbrace{e^{-4\alpha^2}{}_1F_1^2(-n,1;4\alpha^2)}_{\leq \delta^2},
\end{equation}
which exponentially tends to $1$ and it is lower bounded by $S \geq 1 - \frac{1}{2}\delta^2$ in the pDSC region.


\begin{thebibliography}{99}
\bibitem{fidBS} We numerically found that the fidelity between the BS eigenstates and the Rabi eigenstates is greater than $99\%$ before the Juddian points.
\end{thebibliography}
\end{document}